\documentclass[twocolumn,superscriptaddress,preprintnumbers,aps,amsmath,amssymb]{revtex4}

\usepackage[T1]{fontenc}
\usepackage[latin1]{inputenc}
\usepackage[usenames,dvipsnames]{color}

\usepackage{rotating}
\usepackage{bm}        
\usepackage{amssymb}   
\usepackage{amsmath} 
\usepackage{bm}
\usepackage{CJK}

\def\k1{k_1}
\def\k2{k_2}
\def\q1{q_1}
\def\q2{q_2}

\def\({\left (}
\def\){\right )}
\def\[{\left [}
\def\]{\right ]}

\newcommand{\beq}{\begin{equation}}
\newcommand{\eeq}{\end{equation}}

\begin{document}
\date{\today}
\flushbottom \draft
\title{Magnetic tuning of ultracold barrierless chemical reactions} 

\author{Timur V. Tscherbul}
\affiliation{Department of Physics, University of Nevada, Reno, Nevada 89557, USA}
\author{Jacek K{\l}os}
\affiliation{Department of Chemistry and Biochemistry, University of Maryland College Park, College Park, Maryland, 20742, USA}

\begin{abstract}

While attaining external field control of bimolecular chemical reactions has long been a coveted goal of physics and chemistry, the role of hyperfine interactions and dc magnetic fields in achieving such control has remained elusive. 
 We develop an extended coupled-channel statistical theory of barrierless atom-diatom chemical reactions, and apply it to elucidate the effects of magnetic fields and hyperfine interactions on the ultracold chemical reaction Li($^2\text{S}_{1/2}$)~+~CaH($^2\Sigma^+$)~$\to$~LiH($^1\Sigma^+$)~+~Ca($^1\text{S}_{0}$)  on a newly developed set of {\it ab initio} potential energy surfaces. We observe large field effects on the reaction cross sections, opening up the possibility of controlling ultracold barrierless chemical reactions by tuning selected hyperfine states of the reactants with an external magnetic field.

\end{abstract}

\maketitle
\clearpage
\newpage


Using external electromagnetic fields to control chemical reactivity is a central goal of chemical physics \cite{Zare:98,Shapiro:12}, which stimulated the development of new research avenues ranging from mode-selective chemistry \cite{Zare:98} and coherent control \cite{Shapiro:12} to the study of stereodynamics and vector correlations in molecular collisions \cite{Herschbach:06,Perreault:17,Perreault:18} and ultracold controlled chemistry \cite{Krems:08,Balakrishnan:16}. Molecular chemical reactions are most readily controlled  at ultralow temperatures, where the reactants can be prepared in single internal and motional quantum states \cite{Bohn:17}, which maximizes the effects of external electromagnetic fields \cite{Lemeshko:13} and allows for the manifestation of  quantum phenomena, which would otherwise be obscured by thermal averaging, such as threshold and resonance scattering \cite{Klein:16,Bohn:17,Balakrishnan:16}, tunnelling \cite{Balakrishnan:01,Balakrishnan:16}, and interference  \cite{Park:17,Blackmore:18}.
Recent examples include the observation of resonance scattering in low-temperature He$^*$~+~H$_2$  \cite{Klein:16}, He~+~NO  \cite{Vogels:15}, and NO~+~H$_2$ \cite{Vogels:18} collisions, stereodynamical control of low-temperature H$_2$~+~HD collisions in merged molecular beams \cite{Perreault:17,Perreault:18}, and chemical reactions in trapped ensembles of alkali-metal dimers  \cite{Ni:10,Ye:18}, and atom-dimer mixtures \cite{Yang:19}. 
The vast majority of the previous control studies have focused on the rovibrational and nuclear spin degrees of freedom of the reactants.
 In particular,  the chemical reaction KRb + KRb $\to$ K$_2$ + Rb$_2$ can be efficiently suppressed by preparing the reactants in the same rotational and nuclear spin states \cite{Ospelkaus:10} and stimulated by applying an external electric field, which modifies the $p$-wave centrifugal barrier preventing the reaction of two identical fermionic molecules \cite{Ni:10,Miranda:11}.

Recent experimental advances in laser cooling and trapping \cite{Shuman:10,Barry:14} have led to the production of dense, trapped ensembles of molecular radicals ({\it i.e.} molecules with nonzero electron spins) such as CaF($^2\Sigma^+$) \cite{Truppe:17,Anderegg:17, Anderegg:18,Cheuk:18}, SrF($^2\Sigma^+$) \cite{McCarron:18}, YbF($^2\Sigma^+$), \cite{Lim:18} and SrOH($^2\Sigma^+$) \cite{Kozyryev:17}. Cotrapping of these molecules with ultracold alkali-metal atoms \cite{McCarron:18,Lim:18} would open up the fascinating prospect of studying spin-selective ultracold controlled chemistry \cite{Tscherbul:06,Abrahamsson:07,Krems:08}. Specifically,  the electron spins of the reactants can be polarized in an external magnetic field to form a nearly spin-pure  state in the entrance reaction channel corresponding to, {\it e.g.,} the maximum possible total spin $S$ of the reaction complex  \cite{Tscherbul:06,Krems:08}. Because such high-$S$ states are typically non-reactive, the chemical reaction of  spin-aligned reactants [($A(\uparrow$)~+~$B(\uparrow$)] will be suppressed compared to that of spin-antialigned reactants  [($A(\downarrow$)~+~$B(\uparrow$)].  


However, theoretical studies of the effects of spin polarization, hyperfine interactions, and external magnetic fields on atom-molecule chemical reactions have been limited to reactions of weakly bound Feshbach molecules \cite{Knoop:10, Li:19}, save for a recent model study of  ultracold NH-NH reactive scattering  in a magnetic field \cite{Janssen:13}, which did not include the hyperfine structure of NH and focused on collisions of fully spin-polarized molecules.
As a result, the effects of hyperfine interactions and magnetic fields on ultracold reaction dynamics remain  unexplored, limiting our ability to use the fields as a tool to control chemical reactivity at ultralow temperatures.

Here, we develop a theoretical approach to ultracold  reaction dynamics in a magnetic field based on a rigorous  coupled-channel statistical (CCS) model \cite{Rackham:03,Rackham:03a,Alexander:04,GonzalezLezana:07}. The CCS model postulates the existence of a long-lived reaction complex formed temporarily when the reactants get trapped in a long-lived resonance state \cite{Miller:70,Clary:90,Rackham:03a,Alexander:04,Rackham:03}, a powerful idea that forms the basis for quantum threshold models \cite{Quemener:10,GonzalezMartinez:14} and quantum defect theories \cite{Idziaszek:10,Gao:10,Quemener:12}.
The CCS approach rigorously accounts for the multichannel nature of molecular wavefunction in the entrance and exit reaction channels  \cite{Rackham:03,Rackham:03a,Alexander:04} and it 
 has been successfully applied to calculate low-temperature inelastic \cite{Dagdigian:17} and reactive \cite{Quemener:10,Quemener:12,Croft:17,Makrides:15,Tscherbul:15a} collision rates.  Building on the previous work, we extend the CCS approach to explicitly include the effects of hyperfine interactions and external magnetic fields  in the entrance reaction channel, which allows us to explore the magnetic field dependence of the reaction cross sections. We exemplify the extended CCS approach by applying it to the chemical reaction  Li~+~CaH $\to$ LiH~+~Ca on  a newly developed set of {\it ab initio} potential energy surfaces (PESs).  Our field-free results are in good agreement with experiment at $T=1$~K~\cite{Singh:12}. We find that the reaction can be efficiently suppressed  by tuning the  hyperfine states of the reactants with an external magnetic field, opening up the possibility for controlling ultracold spin-dependent chemical reactions. Minimizing atom-molecule reaction rates is essential for efficient sympathetic cooling, in which molecules are immersed in a gas of ultracold atoms and refrigerated by elastic collisions \cite{Carr:09,Tscherbul:11,Lim:15,Morita:18}. Our results thus show that sympathetic cooling of chemically reactive $^2\Sigma$ radicals  could be facilitated by applying external magnetic fields. 


{\it Theory.}  The original CCS theory  \cite{Rackham:03,Rackham:03a}  relates the state-to-state reaction probability $P_{\gamma_A\gamma_B\to \gamma_A'\gamma_B'}$ to the capture probabilities in the entrance and exit reaction channels  $p_{\gamma_A\gamma_B}$ and  $p_{\gamma_A'\gamma_B' }$ as
$P_{\gamma_A\gamma_B\to \gamma_A'\gamma_B'} = {p_{\gamma_A\gamma_B} p_{\gamma_A'\gamma_B'} } /\mathcal{N}$,
where $\gamma_A$ and $\gamma_B$ refer to the incident rovibrational  and hyperfine  states of the reactants (molecule $A$ and atom $B$), and $\mathcal{N}={ \sum_{\gamma_A'\gamma_B'} p_{\gamma_A'\gamma_B' }}$ is a normalization factor.
To obtain the capture probabilities $p_{\gamma_A\gamma_B}$, we solve the Schr\"odinger equation for the atom-molecule reaction complex described by the Hamiltonian (in atomic units, where $\hbar=1$) \cite{Rackham:03,Rackham:03a,Alexander:04,Tscherbul:10}
 \begin{equation}\label{Hamiltonian}
\hat{H} = -\frac{1}{2\mu R}\frac{\partial^2}{\partial R^2}R + \frac{\hat{L}^2}{2\mu R^2} + \hat{V}(\bm{R},\bm{r}) + \hat{H}_{A} + \hat{H}_{B} 
\end{equation}
 subject to capture boundary conditions \cite{Rackham:03,Rackham:03a,Alexander:04} as described in the Supplemental Material \cite{SM}. Here, $\bm{R}$ is the atom-molecule separation vector, $\bm{r}$ joins the nuclei in the diatomic molecule, and $\mu$ and $\hat{L}$ are the reduced mass and orbital angular momentum of the collision complex. The asymptotic Hamiltonians  $\hat{H}_A$ and $\hat{H}_B$ account for the rotational, fine, and hyperfine structure of the reactants in the presence of an external magnetic field \cite{SM}, which is crucial for controlling ultracold reaction dynamics, as shown below.
  
The atom-molecule interaction operator  in Eq.~(\ref{Hamiltonian}) is given by $\hat{V}(\bm{R},\bm{r})=\sum_{S,M_S}V_S(R,r,\theta)|SM_S\rangle \langle SM_S|$, where  $V_S(R,r,\theta)$ are the adiabatic  atom-molecule PESs in the entrance reaction channel calculated {\it ab initio} as described in the Supplemental Material \cite{SM},  $S$ is the total spin of the reaction complex and $M_S$ is the projection of $S$ on the magnetic field axis. Figure 1(a) shows that   both the singlet and triplet PES are strongly anisotropic. The global minimum of the singlet PES is about twice  as deep as that of the  triplet  PES. The approach of Li from the Ca side of CaH ($\theta=180^\text{o}$)  is much more energetically favorable on the singlet PES, which has a deep local minimum in the linear configuration.

   \begin{figure}[t]
\centering
	\includegraphics[width=0.48\textwidth, trim = 5 20 0 0]{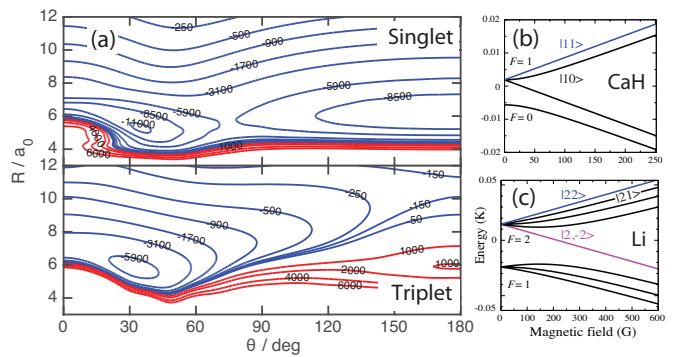}
	\renewcommand{\figurename}{Fig.}
	\caption{(a) Contour plots of the {\it ab initio} Li-CaH PES of singlet ($S=0$, top) and triplet ($S=1$, bottom) symmetries. (b), (c) Zeeman energy levels of CaH and Li. The initial hyperfine states   used in our CCS calculations are labeled as $|F_i m_{F_i}\rangle$ ($i=A,B$) [$F_i$ is an approximate quantum number at $B>0$].}\label{fig:pes}
\end{figure}

To solve the quantum reactive scattering problem in the presence of an external magnetic field, we expand the eigenfunctions of the Hamiltonian  (\ref{Hamiltonian}) 
in eigenstates of the total angular momentum of the reaction complex $|JM\Omega\rangle$ multiplied by the eigenstates of the fragment Hamiltonians $\hat{H}_A$ and $\hat{H}_B$ \cite{SM}. The resulting coupled-channel (CC) equations are solved  numerically  \cite{SM} by  initializing  the complex multichannel log-derivative matrix $\bm{Y}$ \cite{Alexander:04} at the capture radius $R_c$ corresponding to the formation of the reaction complex, 
$\bm{Y}(R_c) = \bm{C}(R_c) \bm{Y}_d(R_c)\bm{C}^T(R_c)$. Here,
  $\bm{C}(R_c)$ are the eigenvectors of the potential coupling matrix, and  $\bm{Y}_d(R_c)$  is the diagonal eigenvalue matrix initialized using the Airy boundary conditions \cite{SM}. 
Having specified the initial value of $\bm{Y}$, we propagate the log-derivative matrix out to a large value of $R$ in the asymptotic region.
The matrix elements  of the Hamiltonian (\ref{Hamiltonian}) are evaluated as described in our previous work \cite{Tscherbul:10} with the following essential modifications: (1) both the singlet and triplet PES of Li-CaH are included in CCS calculations \cite{SM}; (2) the singlet PES is modified at $R=R_m$ to account for its reactive nature
 (the results of the calculations are largely insensitive to $R_m$ \cite{SM}); (3) the hyperfine degrees of freedom of the reactants are explicitly included, as are their interactions with an external magnetic field \cite{SM}.
The final outcome of the calculations is the scattering $S$-matrix, which defines the reaction and capture probabilities \cite{Rackham:03,Rackham:03a,SM}.

\begin{figure}[t]
\centering
	\includegraphics[width=0.4\textwidth, trim = 5 15 0 -20]{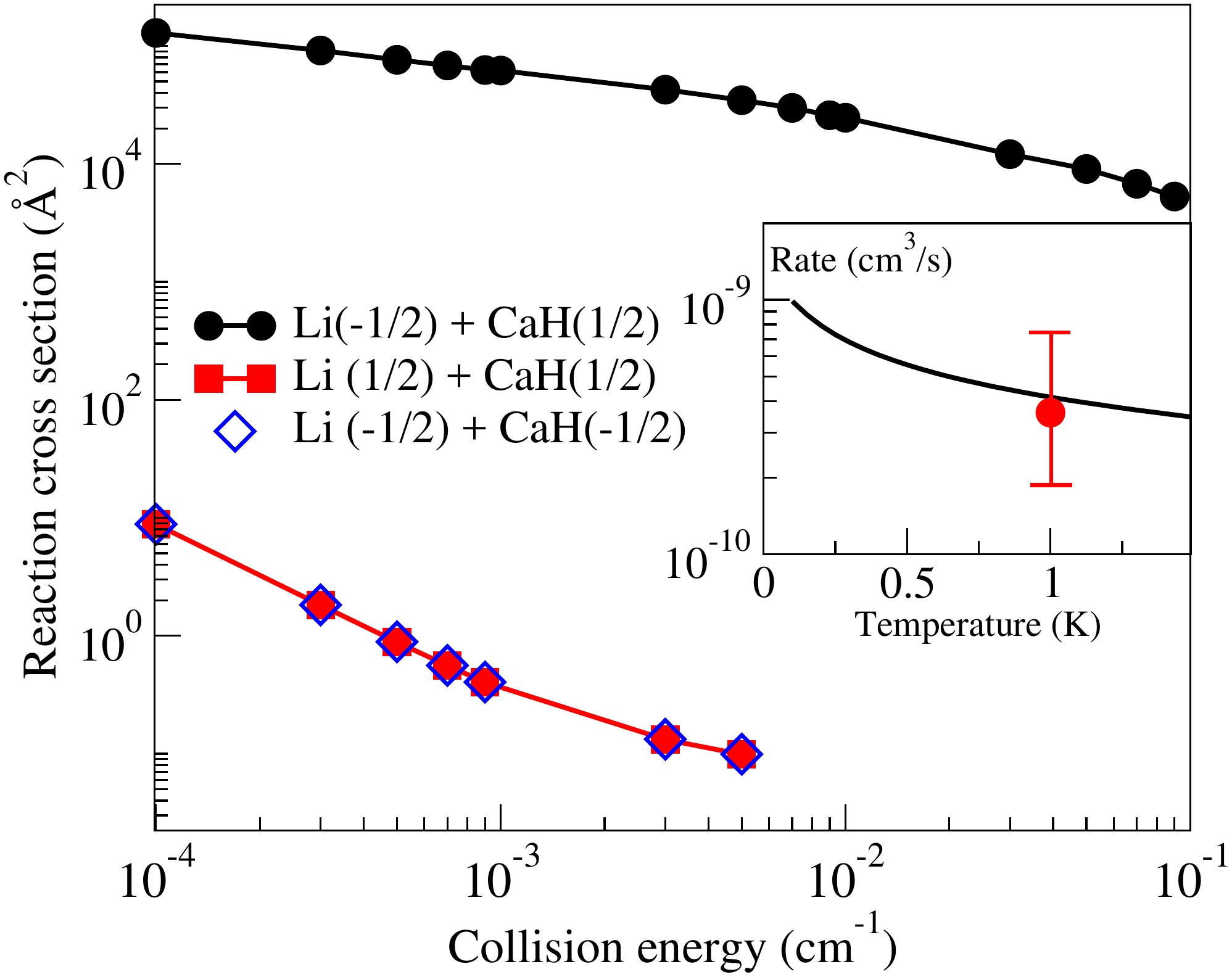}
	\renewcommand{\figurename}{Fig.}
	\caption{Total reaction cross sections for $|\frac{1}{2},\frac{1}{2}\rangle_\text{Li}$ + $|\frac{1}{2},\frac{1}{2}\rangle_\text{CaH}$ (squares) and  $|\frac{1}{2},-\frac{1}{2}\rangle_\text{Li}$ + $|\frac{1}{2},\frac{1}{2}\rangle_\text{CaH}$ (circles)  as a function of the  collision energy at a magnetic field of 0.01 T. Inset: Li+CaH reaction rate plotted as a function of temperature (solid line) vs. experimental result \cite{Singh:12} (circle).}\label{fig:reaction_xs_nohf}
\end{figure}

{\it Results.} We now apply the extended CCS methodology to explore the effect of tuning the Zeeman states of the reactants on the ultracold chemical reaction Li~+~CaH $\to$ LiH~+~Ca.  Figure~\ref{fig:reaction_xs_nohf} shows the collision energy dependence of the  reaction cross section calculated for the different spin states of the reactants $|Sm_S\rangle$ with the hyperfine structure omitted for the moment.  The reaction cross section for  spin-antialigned  reactants $|\frac{1}{2},-\frac{1}{2}\rangle_\text{Li}$~+~$|\frac{1}{2},\frac{1}{2}\rangle_\text{CaH}$ decreases with the collision energy $E_C$ as expected for the Langevin cross section ($\sigma^R\simeq E^{-1/3}$ \cite{Groenenboom:10}). By averaging the dependence $\sigma^R(E)$ over a Maxwell-Boltzmann distributions of collision energies, we obtain the reaction rate  in quantitative agreement with the measured value of $ 3.6\times 10^{-10}$ cm$^3$/s \cite{Singh:12}.

As shown in Fig.~\ref{fig:reaction_xs_nohf}, the reaction of spin-aligned reactants $|\frac{1}{2},\frac{1}{2}\rangle_\text{Li}$~+~$|\frac{1}{2},\frac{1}{2}\rangle_\text{CaH}$ is suppressed by four orders of magnitude compared to that of spin-antinaligned initial states. The spin-aligned reaction nevertheless occurs through the intramolecular spin-rotation and  intermolecular  magnetic dipole-dipole interactions, which flip  the total spin of the Li-CaH complex in the entrance reaction channel \cite{Tscherbul:06,Abrahamsson:07}. Because these interactions are weak, the spin-aligned reaction rate is small, and is comparable to that of non-reactive spin relaxation in $|\frac{1}{2},\frac{1}{2}\rangle_\text{Li}$~+~$|\frac{1}{2},\frac{1}{2}\rangle_\text{CaH}$ collisions \cite{Tscherbul:11}.  

\begin{figure}[t]
\centering
	\includegraphics[width=0.4\textwidth, trim = 5 0 0 -20]{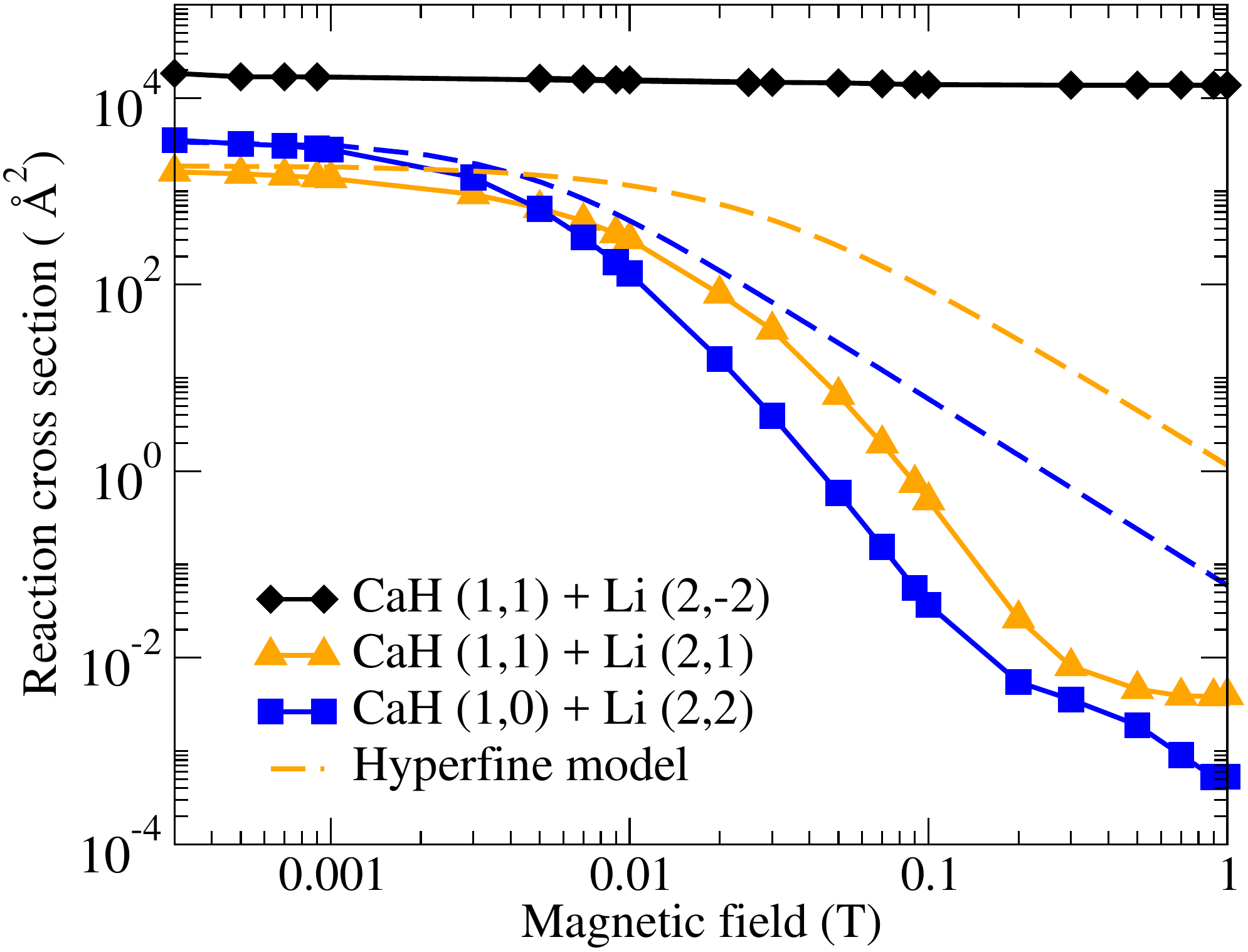}
	\renewcommand{\figurename}{Fig.}
	\caption{(a) Total reaction cross sections for  $|F_A m_{F_A}\rangle_\text{CaH}$ + $|F_B m_{F_B}\rangle_\text{Li}$ as a function of the applied magnetic field. The collision energy is  $10^{-3}$~cm$^{-1}$ = 1.4 mK.  The dashed lines show the predictions of the hyperfine model.}\label{fig:reaction_xs}
\end{figure}

We next explore the effects of external magnetic fields and hyperfine interactions on chemical reactivity.  Figure~\ref{fig:reaction_xs} shows the magnetic field dependence of reaction cross sections for the different initial hyperfine states of Li and CaH [see Figs.~\ref{fig:pes}(b)-(c)]. We observe that certain combinations of initial hyperfine states  are far more reactive than others: In particular,  changing the initial state from  $|11\rangle_\text{CaH} + |2,1\rangle_\text{Li}$ to $|11\rangle_\text{CaH} + |2,-2\rangle_\text{Li}$ enhances the reaction by a factor of 5.
This suggests the possibility of controlling ultracold reaction rates by tuning the hyperfine states of the reactants, which  could be  realized experimentally via radiofrequency and/or optical pumping.


Remarkably, as shown in Fig.~\ref{fig:reaction_xs}, the reactivities of selected initial hyperfine states are extremely sensitive to the magnetic field strength, which opens up the prospect of controlling ultracold barrierless chemical  reactions with external magnetic fields.
To gain insight into the extreme field dependence of the reaction cross sections, we observe that the nuclear spin degrees of freedom (DOF) do not directly participate in the reaction dynamics, which is  governed instead by the spin DOF. This implies, in the spirit of the degenerate internal states approximation \cite{Moerdijk:96,Sikorsky:18},  that the matrix elements of the atom-molecule PES are diagonal in the nuclear spin projections  $m_{I_A}$ and $m_{I_B}$.   
The reaction cross sections
are given in terms of the  exact $S$-matrix elements
\begin{equation}\label{Tmatrix}
\sigma^R_{\gamma_A m_A \gamma_B m_B \to f} = \frac{\pi}{k_i^2} \sum_M | S^M_{\gamma_Am_A \gamma_B m_B \to f}  |^2
\end{equation}
The initial Zeeman states $|\gamma_i m_i\rangle$  ($i=A,\,B$) are linear combinations of the electron and nuclear spin states 
$| \gamma_i m_i\rangle = \sum_{m_{S_i}}  C_{m_{S_i}m_{I_i},\gamma_i}(B) |S_i m_{S_i}\rangle |I_i m_{I_i}\rangle$,
with the  $B$-dependent mixing coefficients $C_{m_{S_i}m_{I_i},\gamma_i}(B)$ \cite{SM} (suppressing the fixed labels $S_i$). Combining this with Eq. (\ref{Tmatrix}), we obtain 
\begin{multline}\label{TmatrixUncoupled}
S^M_{\gamma_Am_A \gamma_B m_B\to f} = \sum_{m_{S_A},m_{S_B}}C_{m_{S_A}m_{I_A},\gamma_A m_A}(B) \\ \times  C_{m_{S_B}m_{I_B},\gamma_B m_B}  (B) S^M_{m_{S_A}  m_{S_B} \to f}  
\end{multline}
We assume that the $S$-matrix elements on the right are independent of $B$ (which is approximately true as shown in Fig.~\ref{fig:reaction_xs}) and they are different from zero only if $m_{S_A}$ and $m_{S_B}$ correspond to the reactive singlet state ($S=0)$ as discussed above.   The magnetic field dependence of the reaction cross section is thus encapsulated in  the hyperfine mixing coefficients $C_{m_{S_i}m_{I_i},\gamma_im_i}(B)$.

We now illustrate the hyperfine model (\ref{TmatrixUncoupled}) by applying it to the chemical reaction $|11\rangle_\text{CaH}$ + $|21\rangle_\text{Li}$.
From Eq. (\ref{TmatrixUncoupled}), we obtain the hyperfine $S$-matrix element as $S^M_{11,21\to f}  =   C_{-\frac{1}{2}\frac{3}{2},21}(B) S^M_{\frac{1}{2}, -\frac{1}{2} \to f} $ and hence
\begin{equation}\label{HFmodel}
\sigma^R_{11,21\to f} =  |C_{-\frac{1}{2}\frac{3}{2},21}(B)|^2  \sigma^R_{\frac{1}{2}, -\frac{1}{2} \to f}
\end{equation}
where $\sigma^R_{\frac{1}{2}, -\frac{1}{2} \to f}$ is the reaction cross section in the absence of the hyperfine structure (upper trace in Fig.~\ref{fig:reaction_xs_nohf}).

As shown in Fig.~\ref{fig:reaction_xs}, the reaction cross sections predicted by the hyperfine model decrease  as a function of the applied magnetic field, in qualitative agreement with the CCS results. The suppression is due to the   decoupling of the electron and nuclear spins: As shown in Fig.~\ref{fig:pes}(c), the initial state  $|2,1\rangle_\text{Li}$ correlates with the nonreactive Zeeman state $|m_{S}=1/2\rangle_\text{Li}$  in the high-field limit, leading to a decrease of the contribution of the reactive state $|m_{S}=-\frac{1}{2}\rangle_\text{Li}$. The reaction cross section scales as $B^{-2}$ due to the mixing coefficient $C_{-\frac{1}{2}\frac{3}{2},21}(B)\simeq B^{-1}$. We note that the  hyperfine model predicts a less steep decline of the reaction cross sections with the field. This is expected because the hyperfine model neglects the effects of the magnetic field on the $S$-matrix elements, which are likely to become more pronounced at higher fields \cite{Moerdijk:96,Sikorsky:18}.  

In conclusion, we have extended the rigorous CCS model of barrierless chemical reactions \cite{Rackham:03,Rackham:03a,Alexander:04} to include the hyperfine structure of open-shell reactants and their interactions with external magnetic fields. 
 We have applied the model to explore the effects of hyperfine interactions and magnetic fields on the dynamics of the prototypical barrierless chemical reaction CaH~+~Li $\to$ LiH + Ca.
Our calculated reaction rates agree with  experiment \cite{Singh:12} and display a dramatic dependence on the external magnetic field, which could be used to facilitate  sympathetic cooling of chemically reactive $^2\Sigma$ molecules with alkali-metal atoms \cite{Tscherbul:11,Warehime:15}. We expect our approach to be readily applicable to a wide range of ultracold barrierless chemical reactions of current experimental interest,  including those involving molecular ions \cite{Puri:17,Yang:18,Zhang:17,Hall:12} and alkaline-earth halides SrF and CaF \cite{McCarron:18,Morita:18,Meyer:11}.

We are grateful to Gerrit Groenenboom, Roman Krems, and Masato Morita for encouraging discussions.  This work was supported by NSF grant No. PHY-1607610. J.~K. acknowledges financial support under the U. S. National Science Foundation grant No. CHE-156872 to M.~H. Alexander.

\clearpage
\newpage

\section*{Supplemental Material for the manuscript  ``Magnetic tuning of ultracold barrierless 
chemical reactions''}


\setcounter{figure}{0}

\vspace{-3mm}

This Supplemental Material provides a brief overview of the extended coupled-channel statistical (CCS) theory of barrierless atom-diatom chemical reactions [Sec. I] along with the technical details of our {\it ab initio} calculations of the Li-CaH potential energy surfaces (PES) [Sec. II].  The details of numerical calculations and convergence tests follow in Sec.~III.

\vspace{-2mm}

\section{The extended CCS model: Overview and Numerical implementation }  

This section provides an overview of the extended CCS approach. We begin by  introducing the CC equations in Sec.~IA and describing the procedure of applying the boundary conditions  in Sec. IB.
 Sec. IC describes further technical details pertaining to  the evaluation of the matrix elements of the atom-molecule interaction and of the orbital angular momentum of the collision complex.  

\subsection{ Numerical solution of CC equations: Reaction cross sections and  capture probabilities}

The CCS  capture probability in the entrance reaction channel is given by \cite{Rackham:03,Rackham:03a,Alexander:04}
 \begin{equation}\label{pcapture}
p^M_{\gamma_A\gamma_B l} = 1 - \sum_{\gamma_A'\gamma_B'l'} |  S^M_{\gamma_A'\gamma_B' l',\gamma_A\gamma_Bl}|^2
\end{equation}
where $\gamma_A$, $\gamma_B$  and $\gamma_A'$, $\gamma_B'$  stand for the initial and final  Zeeman states of the reactants, $l$ and $l'$ are the corresponding orbital angular momenta, and $M$ is the space-fixed (SF) projection of the total angular momentum $J$ of the collision complex on the magnetic field axis, which is conserved for reactions in magnetic fields. 
The total reaction cross section is obtained  by summing the entrance channel capture probabilities (\ref{pcapture}) over  a range of orbital angular momenta $l$ and total angular momentum projections $M$ 
 \begin{equation}\label{sigmaR}
\sigma_{\gamma_A\gamma_B \to f} = \frac{\pi}{k_{\gamma_A\gamma_B}^2}  \sum_M \sum_l    p_{\gamma_A\gamma_B l}^M
\end{equation}
where $k_{\gamma_A\gamma_B}=2\mu  E_C$ is the wavevector in the incident collision channel and $E_C$ is the collision energy.
We note that the reaction cross section can be obtained from the fully state-to-state cross section by summing over  the final LiH~+~Ca product states $\gamma_A'$ and $\gamma_B'$.

The $S$-matrix elements in Eq.~(\ref{pcapture}) are obtained from the  radial solutions $F^{M}_{\alpha_A\alpha_B J\Omega}(R) $ of the  coupled-channel (CC) equations at total energy $E$ \cite{Rackham:03,Rackham:03a,Alexander:04}
\begin{multline}\label{CC}
\left[\frac{d^2}{dR^2} + 2\mu E\right] F^{M}_{\alpha_A\alpha_B J\Omega}(R) \\ = 2\mu \sum_{\alpha_A'\alpha_B'} \sum_{J',\Omega'} \langle \alpha_A\alpha_B|\langle JM\Omega |\hat{V}(R,r,\theta)  + \frac{\hat{L}^2}{2\mu R^2} + \hat{V}(R,r,\theta) \\ + \hat{H}_\text{as}| \alpha_A'\alpha_B'\rangle  |J'M\Omega'\rangle  F^M_{\alpha_A'\alpha_B'J'\Omega'}(R).
\end{multline}
subject to the capture boundary conditions as described below.  The CC equations (\ref{CC}) describe   atom-molecule scattering in the entrance reaction channel in the presence of an external magnetic field.  In Eq. (\ref{CC})
 \begin{multline}\label{basis}
 |\alpha_A\alpha_B\rangle |JM\Omega\rangle = |\alpha_A\rangle |\alpha_B\rangle |JM\Omega\rangle \\ = |N_A K_{N_A}\rangle |S_A\Sigma_A\rangle  \times |I_A\Sigma_{I_A}\rangle |S_B\Sigma_B\rangle |I_B\Sigma_{I_B}\rangle |JM\Omega\rangle 
 \end{multline}
 are body-fixed (BF) basis functions for the overall rotational motion ($|JM\Omega\rangle$) and the internal degrees of freedom of molecule $A$ ($\alpha_A$) and atom $B$ ($\alpha_B$), including   the rotational angular momentum $N_A$, the electron spins $\hat{S}_A$ and $\hat{S}_B$ and the nuclear spins $\hat{I}_A$ and $\hat{I}_B$, with  $K_{N_A}$, $\Sigma_A$, $\Sigma_B$, $\Sigma_{I_A}$, and $\Sigma_{I_B}$ being the projections of $N_A$, $S_A$, $S_B$, $I_A$, and $I_B$ on the atom-diatom separation vector $\mathbf{R}$ chosen as the $z$-axis of the BF coordinate frame \cite{Tscherbul:10}. The matrix elements are evaluated as described in Sec.~IC below.

 \begin{figure}[t]
\centering
	\includegraphics[width=0.4\textwidth, trim = 0 0 0 0]{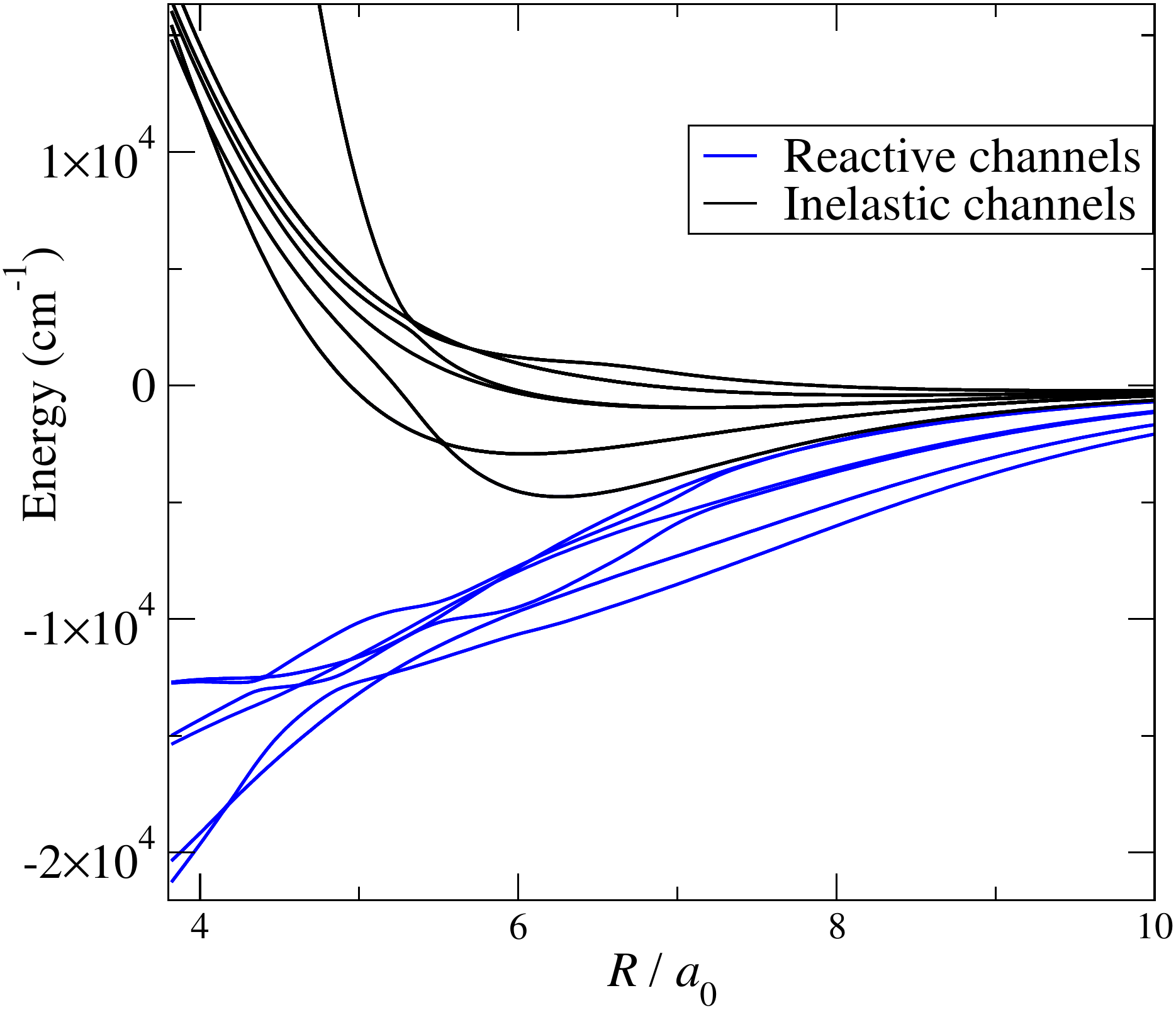}
	\renewcommand{\figurename}{Fig.}
	\caption{Reactive (blue/grey lines) and inelastic (black lines) adiabatic channels calculated for Li-CaH using a restricted basis set with $N_\text{max}=2$ and total angular momentum projection $M=0$ at a magnetic field of 0.01 T.  Note the different asymptotic behavior of the inelastic and reactive channels at short $R$. }\label{fig:S1}
\end{figure}

 \subsection{Boundary conditions}
We solve the CC equations numerically by constructing the log-derivative matrix $\mathbf{Y}=\bm{\Psi}^{-1}\bm{\Psi}$, where $\Psi$ is the wavefunction matrix, and propagating it from a small value of $R=R_c$ out to the asymptotic region \cite{Rackham:03,Rackham:03a,Alexander:04}.  We  choose an initial value of the capture radius $R=R_c$ inside the reaction complex region and initialize the complex symmetric log-derivative matrix as \cite{Rackham:03,Rackham:03a,Alexander:04}
\begin{equation}\label{YRc}
\mathbb{Y}(R_c) = \mathbb{C}(R_c) \mathbb{Y}_d(R_c) \mathbb{C}^T(R_c)
\end{equation}
where $\mathbb{Y}_d(R_c)=(y_1(R_c),y_2(R_c),\ldots,y_N(R_c))$ is the diagonal matrix constructed from the  eigenvalues $\epsilon_c=-k_c^2$ of the coupling matrix $\mathbf{W}(R_c)$ 
\begin{multline}\label{W}
[\mathbf{W}(R_c)]_{nn'} = \langle JM\Omega | \langle \alpha_A \alpha_B| V(R_c,r,\theta) + \frac{L^2}{2\mu R_c^2}  \\+ \hat{H}_\text{as}| \alpha_A' \alpha_B' \rangle |J'M'\Omega'\rangle
\end{multline}
using the multichannel Wentzel-Kramers-Brillouin (WKB)  boundary conditions \cite{Alexander:04}. The entrance channels of chemical reactions that occur on multiple PESs (such as the Li-CaH reaction considered here) typically include  the highly attractive as well as strongly repulsive PESs, leading to two qualitatively different types of adiabatic channels illustrated in Fig 1.  The  reactive channels  decrease in energy with decreasing $R<R_c$, whereas the  nonreactive channels show the opposite trend.   
Both types of channels can be treated on an equal footing using the Airy boundary conditions \cite{Alexander:04,Manolopoulos:06}. Following Ref.~\cite{Manolopoulos:06}, we initialize the  elements of   the diagonal matrix $\mathbb{Y}_d$ as
\begin{equation}\label{yRc}
y(R_c) = W'(R_c)^{1/3}\frac{\phi'(x_c)}{\phi(x_c)}
\end{equation}
where $W'(R_c)^{1/3}$ is the real root of $x^3=W'(R_c)$, and  $W'(R_c)$ is the derivative of the adiabatic eigenvalue  $W(R_c)$ (an eigenvalue of matrix  $\mathbb{W}$), which is positive (negative) for reactive (inelastic) channels [see Fig.~1]. The scaled wavefunction $\phi(x)$ is a solution of the Airy equation  for the adiabatic channels linearly extrapolated into the reaction complex region $R<R_c$ \cite{Manolopoulos:06}
\begin{equation}\label{Airy_eqn} 
\phi''(x) = \frac{a^3}{W'(R_c)^2}x {\phi(x)}
\end{equation}
where $x=W(R)/[W'(R)^{2/3}]$ is a scaled radial coordinate.
 
 For $W'(R_c)>0$ (reactive channels) and  $R\ll R_c$,  we have $x\to -\infty$ and the wavefunction ratio in Eq. (\ref{yRc}) takes the form  \cite{Manolopoulos:06}
\begin{equation}\label{Airy_reac} 
\frac{\phi'(x_c)}{\phi(x_c)} = \frac{\text{Bi}'(x) + i \text{Ai}'(x)}{\text{Bi}(x) + i \text{Ai}(x)} \stackrel{x\to-\infty}\longrightarrow -i\sqrt{-x}-\frac{1}{4x}
\end{equation}
where $\text{Bi}(x)$ and $\text{Ai}(x)$ are the Airy functions, which oscillate in the limit of large negative $x$.

 For $W'(R_c)<0$ (inelastic channels) and  $R\ll R_c$,  we have $x\to +\infty$. Retaining only the asymptotically decaying Airy function $\text{Ai}(x)$, we obtain the wavefunction ratio in Eq. (\ref{yRc})  as \cite{Manolopoulos:06}
\begin{equation}\label{Airy_inel} 
\frac{\phi'(x_c)}{\phi(x_c)} = \frac{\text{Ai}'(x)}{\text{Ai}(x)} \stackrel{x\to+\infty}\longrightarrow \sqrt{x}-\frac{1}{4x}
\end{equation}

In practice, the asymptotic expressions (\ref{Airy_reac})-(\ref{Airy_inel}) give sufficiently accurate results for $|x|\ge 5$. However,  in our numerical calculations a small fraction of adiabatic channels has $|x| < 5$, making it necessary to apply numerically exact expressions for the Airy functions.

At a large atom-molecule distance $R=R_a$, we match the log-derivative matrix $\mathbb{Y}(R_a)$ to the standard incoming and outgoing wave boundary conditions to obtain the scattering $S$-matrix  \cite{Rackham:03,Rackham:03a}
\begin{equation}\label{Airy_inel2} 
\mathbb{S} =  [\mathbb{Y}(R_a)  \mathbb{O}_E(R_a)  - \mathbb{O}_E'(R_a)]^{-1}[ \mathbb{Y}(R_a)  \mathbb{I}_E(R_a)  - \mathbb{I}_E'(R_a) ]
\end{equation}
where $\mathbb{I}_E$ and $\mathbb{O}_E$ are the diagonal matrices composed of the incoming and outgoing-wave solutions of  CC equations in the absence of the atom-molecule interaction (for open channels)
 \begin{align}\label{IOsolutions_open} \notag
[{I}_E(R_a)]_{\gamma l,\gamma'l'}  &= \delta_{\gamma\gamma'}  \delta_{ll'} k_{\gamma}^{1/2}R h_{l}^{(2)}(k_\gamma R) \\
[{O}_E(R_a)]_{\gamma l,\gamma'l'}  &= \delta_{\gamma\gamma'} \delta_{ll'} k_{\gamma}^{1/2}R h_{l}^{(1)}(k_\gamma R) 
\end{align}
 where $\gamma$ is a compound index for $\gamma_A,\gamma_B$, $k_\gamma =[2\mu E_C]^{1/2} $ is the incident wavevector, $E_C$ is the collision energy,  and $h^{(\pm)}_l(x)$ are the spherical Hankel functions.
The asymptotic solutions for closed channels are given by \cite{Rackham:03a}
  \begin{align}\label{IOsolutions_closed} \notag
[{I}_E(R_a)]_{\gamma l,\gamma'l'}  &= \delta_{\gamma\gamma'}  \delta_{ll'} |k_{\gamma}|^{1/2}R i_{l}(|k_\gamma| R) \\
[{O}_E(R_a)]_{\gamma l,\gamma'l'}  &= \delta_{\gamma\gamma'} \delta_{ll'} |k_{\gamma}|^{1/2}R k_{l}(|k_\gamma |R) 
\end{align}
 where $i_{l}(x)$ and $k_{l}(x)$ are the modified spherical Bessel functions.

  \subsection{Matrix elements}

We now turn to the technical details of the evaluation of the matrix elements in the CC equations (\ref{CC}). The asymptotic Hamiltonian may be written as  \cite{Tscherbul:10}
\begin{equation}\label{Has_separated} 
\hat{H}_\text{as}=\hat{H}_A + \hat{H}_B,
\end{equation}
where $\hat{H}_A$ and  $\hat{H}_B$ are the asymptotic Hamiltonians of the reactants [molecule $A(^2\Sigma$) plus  atom $B(^2$S)]. The molecular Hamiltonian is given by \cite{Tscherbul:07,Tscherbul:18b}
\begin{multline}\label{HA} 
 \hat{H}_A = B_e \bm{N}_A^2 + \gamma\hat{\bm{N}}_A\cdot\hat{\bm{S}}_A +  (b+c/3)\hat{\bm{I}}_A\cdot\hat{\bm{S}}_A  + \frac{c\sqrt{6}}{3} \\ \times \biggl{(}\frac{4\pi}{5}\biggr{)}^{1/2}\sum_{q=-2}^2 (-1)^qY_{2-q}(\hat{r}_A)
          [\hat{\bm{I}}_A\otimes \hat{\bm{S}}_A]^{(2)}_q  + 2\mu_0 B S_{A_Z}
\end{multline}
where $\hat{\bm{N}}_A$ is the rotational angular momentum,  $\hat{\bm{S}}_A$ and $\hat{\bm{I}}_A$ are the electron and nuclear spins with space-fixed (SF) projections $\hat{S}_{Z_A}$ and $\hat{I}_{Z_A}$ [$I_A=S_A=1/2$ for CaH($X^2\Sigma$)], $\hat{\bm{I}}_A\otimes \hat{\bm{S}}_A$ is a tensor product of $\hat{\bm{I}}_A$ and $\hat{\bm{S}}_A$,  $Y_{2-q}(\hat{r})$ is a spherical harmonic describing the orientation of the molecular axis $\hat{r}_A$ in the SF frame,  and $B_e$, $\gamma$, $b$, and $c$ are the rotational, spin-rotation, and hyperfine constants. We neglect the weak  nuclear spin-rotation interaction \cite{Tscherbul:07}.
  The atomic Hamiltonian 
 \begin{equation}\label{HB} 
 \hat{H}_B = A_B \hat{\bm{I}}_B\cdot\hat{\bm{S}}_B +2\mu_0 B S_{B_Z}
 \end{equation}
  includes the hyperfine coupling of the electron and nuclear spins parametrized by the atomic hyperfine constant   $A_B$, and the interaction  of the atomic spin with an external magnetic field $B$.
Writing the asymptotic Hamiltonian (\ref{Has_separated}) as a sum of field-free and Zeeman terms
\begin{equation}\label{Has_separated2} 
\hat{H}_\text{as}=\hat{H}_A^{(0)} + \hat{H}_{Z,A} +\hat{H}_B^{(0)} + \hat{H}_{Z,B},
\end{equation}
and taking advantage of the direct-product structure of the BF basis set (\ref{basis}), we obtain
\begin{multline}\label{Has_matrixels} 
\langle \alpha_A \alpha_B | \langle JM\Omega |\hat{H}_\text{as} | \alpha_A' \alpha_B'  \rangle |J'M\Omega' \rangle =  \delta_{JJ'}\delta_{\Omega\Omega'} \\ \times
\left[ \delta_{\alpha_B\alpha_B'} \langle \alpha_A |\hat{H}_A^{(0)}|\alpha_A'\rangle + \delta_{\alpha_A\alpha_A'} \langle \alpha_B |\hat{H}_B^{(0)}|\alpha_B'\rangle  \right] \\ +  \delta_{\alpha_B\alpha_B'} \langle \alpha_A | \langle JM\Omega   |\hat{H}_{Z,A} |  \alpha_A'  \rangle |J'M\Omega' \rangle  \\ 
+ \delta_{\alpha_A\alpha_A'} \langle \alpha_B | \langle JM\Omega   |\hat{H}_{Z,B} |  \alpha_B'  \rangle |J'M\Omega' \rangle
\end{multline}
The matrix elements on the right can be evaluated as described in our previous work  \cite{Tscherbul:07,Tscherbul:10}. Diagonalization of $\hat{H}_\text{as}$ at $B>0$ produces unphysical Zeeman eigenstates, which do not affect low-temperature collision dynamics provided a sufficient number of total angular momentum eigenstates $J_\text{max}$ is included in the CC basis \cite{Tscherbul:10}.


To calculate the matrix elements of the orbital angular momentum operator in Eq.~(\ref{CC}) in the body-fixed angular momentum basis, we express the latter in the form
 \begin{equation}\label{L2}
 \hat{L}^2 = (\hat{J} - \hat{N}_A - \hat{S}_A- \hat{S}_B - \hat{I}_A - \hat{I}_B)^2
 \end{equation}
While  $\hat{L}^2$   can be expressed in terms of the raising and lowering operators for all angular momenta involved as done, {\it e.g.}, in Ref. \cite{Tscherbul:10}, the resulting expressions are rather cumbersome due to the presence of two additional nuclear spin operators $\hat{I}_A$ and $\hat{I}_B$. To simplify the evaluation of the angular momentum matrix elements, it is convenient to define the orbital angular momentum of the atom-molecule system in the absence of the nuclear spin
 \begin{equation}\label{L42}
 \hat{L}_4 = \hat{J} - \hat{N}_A - \hat{S}_A- \hat{S}_B 
 \end{equation}
The matrix elements of this operator can be  evaluated as described in our previous work \cite{Tscherbul:10}. To incorporate the nuclear spins, we combine  Eqs. (\ref{L2}) and (\ref{L42}) and use the fact that the nuclear spin operators commute with $\hat{L}_4$ to obtain
 \begin{multline}\label{L2viaL42}
 \hat{L}^2 = ( \hat{L}_4 - \hat{I}_A- \hat{I}_B )^2 = \hat{L}_4^2 + \hat{I}_A^2 + \hat{I}_B^2 - 2 \hat{L}_4 \cdot \hat{I}_A \\ - 2 \hat{L}_4 \cdot \hat{I}_B  + 2 \hat{I}_A \cdot \hat{I}_B
 \end{multline}
Expressing the scalar products of angular momentum operators via the raising and lowering operators, {\it e.g.},  $\hat{L}_4 \cdot \hat{I}_A=\hat{L}_{4z} \cdot \hat{I}_{Az} + \frac{1}{2}(\hat{L}_{4+} \cdot \hat{I}_{A-}+\hat{L}_{4-} \cdot \hat{I}_{A+})$,  the evaluation of the matrix elements of the operator $\hat{L}^2$ (\ref{L2}) in the basis  (\ref{basis}) reduces to straightforward angular momentum algebra described, {\it e.g.}, in Ref. \cite{Tscherbul:10}.

\begin{figure}[t]
\centering
	\includegraphics[width=0.378\textwidth, trim = 30 0 0 0]{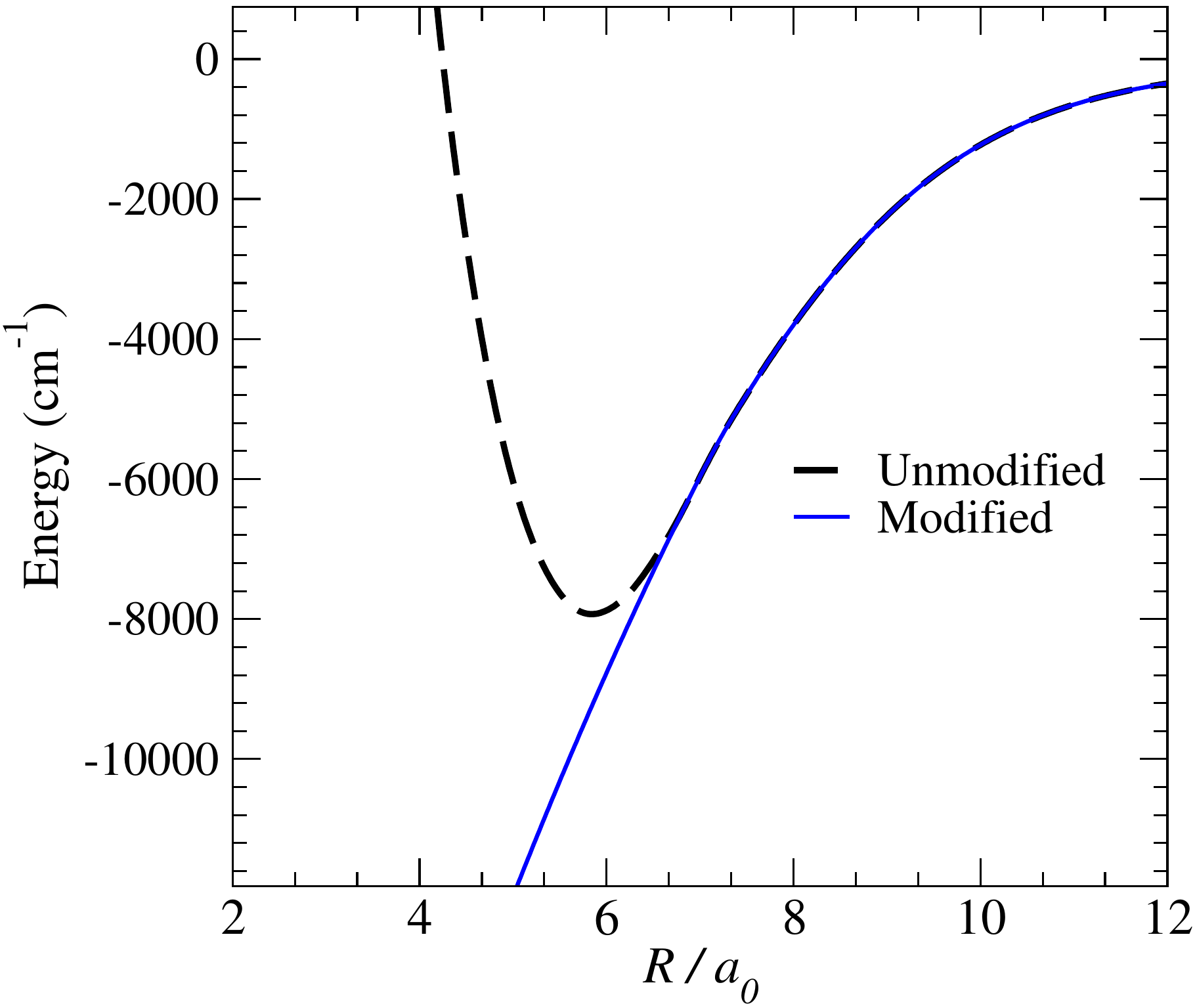}
        \includegraphics[width=0.39\textwidth, trim = 0 0 0 0]{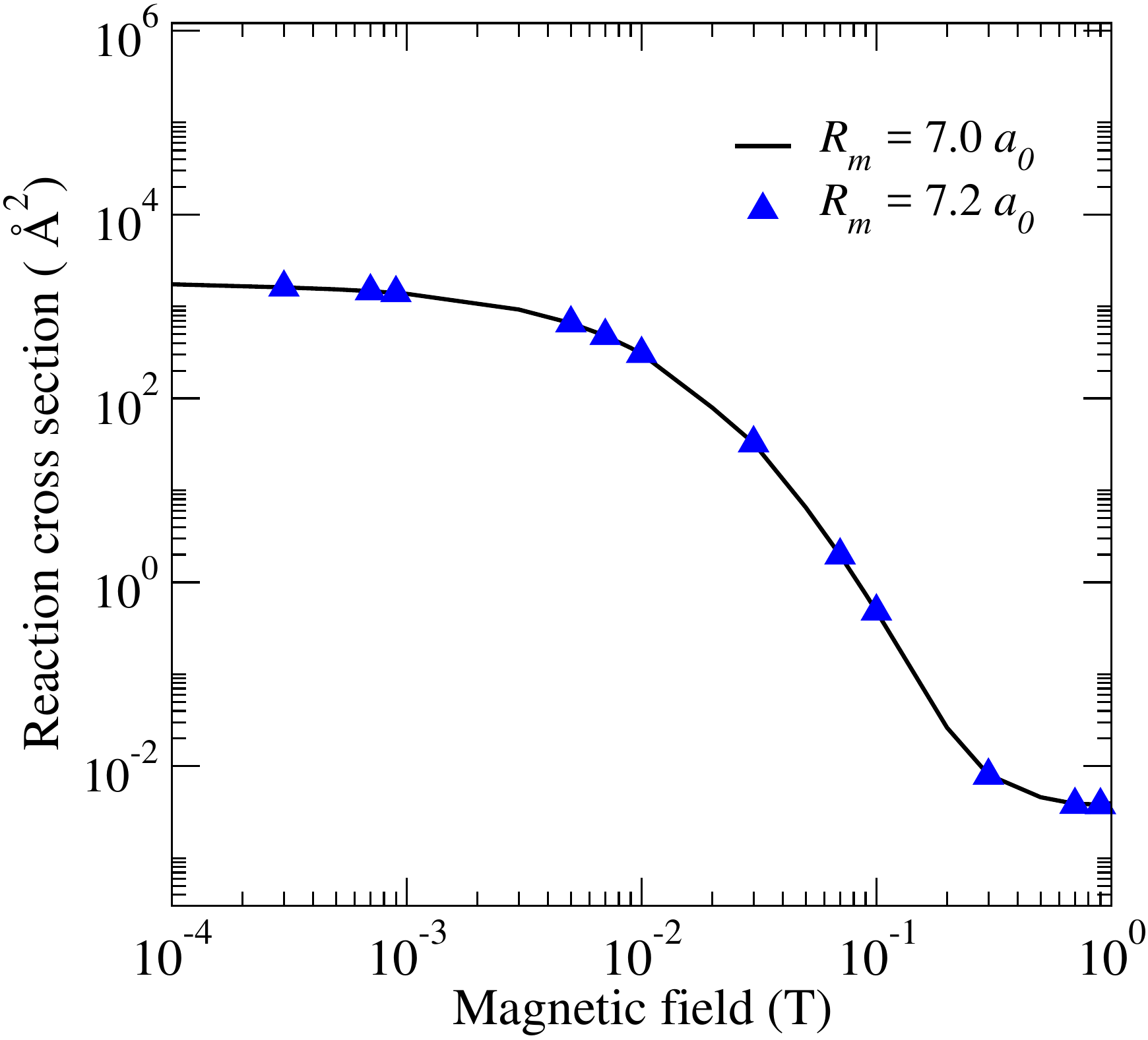}
	\renewcommand{\figurename}{Fig.}
	\caption{(Upper panel): Unmodified (dashed line) and modified (full line) isotropic parts of the singlet Li-CaH PES in the rigid rotor approximation $V_0^{S=0}(R)$. The value of the matching distance $R_m=7\,a_0$. (Lower panel): Magnetic field dependence of the reaction cross section calculated for several values of the matching parameter $R_m$ (in units of $a_0$) and the reactants' initial states $|1,1\rangle_\text{CaH}$ + $|21\rangle_\text{Li}$ at a collision energy of 0.001 cm$^{-1}$.}\label{fig:S2}
\end{figure}

\section{Ab initio calculations and PES fitting}

 To compute the singlet ($S=0$) and triplet ($S=1$) PES of the Li-CaH reaction complex, we used  high-level  multi-reference configuration interaction (MRCI) and coupled-cluster methods  with single, double, and noniterative triple excitations  as implemented in the MOLPRO code  \cite{molpro}.

 The triplet PES  is calculated  as described in our previous work~\cite{Tscherbul:11} at the CCSD(T) level of theory \cite{knowles:93}.  To compute the singlet ($S=0$)  PES, we took into account the multi-reference character of the electronic wavefunction by using the multi-reference configuration interaction (MRCI) method~\cite{werner:88} with single and double excitations (MRCISD) and Davidson corrections (+Q) to approximately account for contributions of higher excitations. The MRCISD+Q calculations were started from the reference orbitals obtained at the state-averaged complete active space self-consistent-field (sa-CASSCF) level treating all $S$ states on the same footing. The active space contained 4 $A'$ and 1 $A''$ orbitals and 6 orbitals were correlated but kept doubly occupied (5 $A'$ and 1 $A''$). The frozen core of the Ca atom was composed of 4 $A'$ and 1 $A''$ orbitals. We used the augmented, correlation consistent triple-zeta  (aug-cc-pvtz) basis for H  \cite{dunning:89}, a quadruple-zeta basis  (aug-cc-pvqz) for Li , and a valence quadruple-zeta (cc-pvqz) basis for  Ca ~\cite{Koput:2002}.  

The {\em ab initio} calculations were performed on a two-dimensional grid of $R$ and $\theta$, with $\theta \in 0^\circ{-}180^\circ$ in steps of 5$^\circ$ and $R\in 3.5{-}30\,a_0$  \cite{molpro}.  
To facilitate the calculation of the matrix elements of the atom-molecule interaction operator $\hat{V}(R,r,\theta)=\sum_{SM_S}V_{S}(R,\theta,r)|SM_S\rangle\langle SM_S|$, where $S$ is the total spin of the reaction complex,  we expand the adiabatic potential energy surfaces $V_{S}(R,\theta,r)$ in Legendre polynomials $P_\lambda(\cos\theta)$
\begin{equation}\label{Vexpansion} 
V_{S} (R,r,\theta) =  \sum_{\lambda=0}^{\lambda_\text{max}} V^S_{\lambda}(R,r)P_\lambda(\cos\theta)
\end{equation}
  Because we neglect the weak nuclear spin-dependent interactions that depend on $R$ (such as the interaction of the nucelar spins with the overall rotation of the reaction complex),  the matrix elements of the interaction potential are diagonal in the nuclear spin quantum numbers $\Sigma_{I_A}$  and  $\Sigma_{I_B}$ [see Eq.~(\ref{basis})]. As a result, the matrix elements of Eq.~(\ref{Vexpansion}) are given by the expressions similar to Eqs.~(30) and (31) of Ref. \cite{Tscherbul:10}.

For use in quantum scattering calculations, the {\em ab initio} data points are expanded in Legendre polynomials (\ref{Vexpansion}) with $\lambda_\text{max}=18$ (for $S=0$) and $\lambda_\text{max}=14$ (for $S=1$). The resulting radial expansion coefficients $V_\lambda(R)$ are fit using the Reproducing Kernel Hilbert Space method of Rabitz and coworkers~\cite{ho:96}. To avoid unphysical distortion of the fit,  we damped the very high repulsive energies at small $R$. 

Following our previous work \cite{Tscherbul:15a}, we invoke the rigid-rotor approximation by freezing  the internuclear distance of CaH at its equilibrium value $r_{e}=2.0025$~\AA. This approximation provides quantitatively accurate capture probabilities for the Li-CaH chemical reaction on a single adiabatic PES  \cite{Tscherbul:15a} at a much reduced computational cost. However, in the context of the CCS model, the rigid-rotor approximation leads to all adiabatic potentials becoming repulsive (i.e. non-reactive) at sufficiently short $R$. To address this, we introduce the following modification of the isotropic part of the singlet (reactive) PES 
 \begin{multline}\label{Rmod}
V^{0}_0(R) = V_0^{0} (R_c) + \frac{dV_0^0(R)}{dR} \biggr\vert_{R_m}(R-R_m)  
\\ + \frac{1}{2} \frac{d^2 V_0^0(R)}{dR^2} \biggr\vert_{R_m}(R-R_m)^2  \qquad (R<R_m)
 \end{multline}
 where $R_m$ is a matching point to the right of the potential minimum, where the first and second derivatives of the potential have opposite signs with the second derivative being negative, so as to ensure the decreasing behavior of Eq. (\ref{Rmod}) with decreasing $R<R_m$ [see Fig.~2(a)].
 The modification replaces the short-range repulsive wall of the rigid-rotor potential with a function that decreases with $R$. This results in a one-parameter family of modified potentials parameterized by the values of $R_{m}$.
 We have verified (see Sec. II below) that the calculated  capture probabilities are  insensitive to the choice of $R_m$, thereby validating the procedure.

\section{Convergence tests}

\begin{figure}[t]
\centering
	\includegraphics[width=0.4\textwidth, trim = 40 0 0 0]{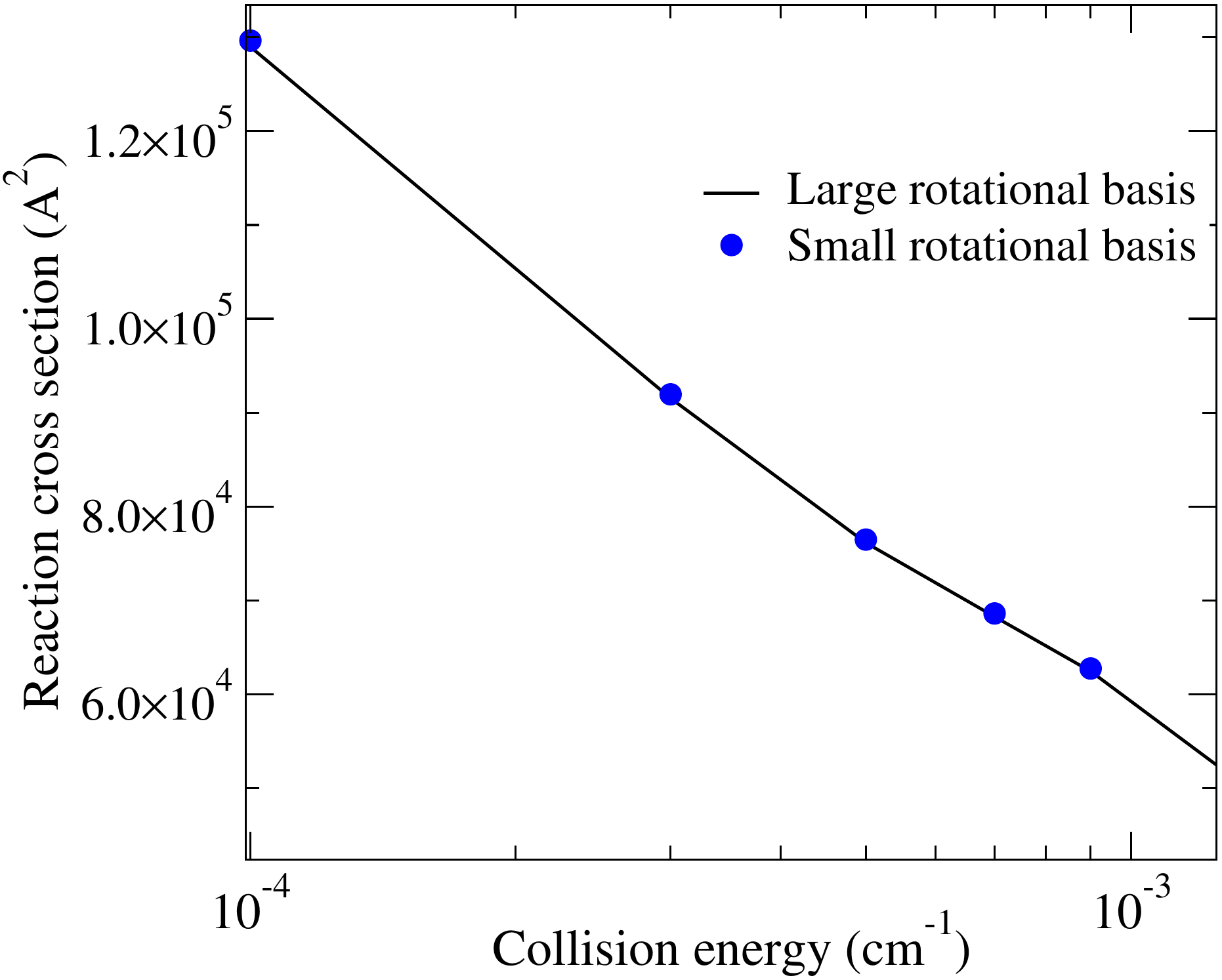}
	\renewcommand{\figurename}{Fig.}
	\caption{Total CCS reaction cross sections calculated using large ($N_\text{max}=55$, black line) and small ($N_\text{max}=2$, circles) rotational basis sets for the initial spin states $|\frac{1}{2},\frac{1}{2}\rangle_\text{CaH}$ + $|\frac{1}{2},-\frac{1}{2}\rangle_\text{Li}$ in the absence of the hyperfine structure. The magnetic field is $B=0.01$~T and $J_\text{max}=2$. }\label{fig:S3}
\end{figure}

We carried out a series of  convergence tests to determine the optimal values of the asymptotic matching distance $R_a$ and the cutoff parameters $N_\text{max}$ and $J_\text{max}$ that determine the  sizes of the rotational and total angular  momentum basis sets.  We use the following values of  $R_a$ to obtain  the capture probabilities and reaction cross sections converged to within 10-20\%: $E_C$: $R_a=360 \,a_0$ ($E_C=10^{-4}-10^{-3}$ cm$^{-1}$), $R_a=270 \,a_0$ ($E_C=10^{-3}-10^{-2}$ cm$^{-1}$), and $R_a=200 \,a_0$ ($E_C>10^{-2}$  cm$^{-1}$), with a uniform  grid step of $0.02\, a_0$. The value of the capture radius $R_c$ was set to $3.84 a_0$ in all of the calculations.

At the lowest collision energies studied in this work ($10^{-4}<E_C<10^{-3}$ cm$^{-1}$) it is sufficient to truncate the total angular momentum basis at $J_\text{max}=2$ to produce results converged to  $<5\%$. At higher collision energies,  progressively higher values of $J_\text{max}$ were used, up to $J_\text{max}=10$ at $E_C=7$ cm$^{-1}$.

We also carried out  convergence tests with respect to the maximum number of rotational states $N_\text{max}$ included in the basis set. The capture cross sections for the spin-antialigned initial states $|\frac{1}{2},\frac{1}{2}\rangle_\text{CaH}$ + $|\frac{1}{2},-\frac{1}{2}\rangle_\text{Li}$ are large and remarkably insensitive to $N_\text{max}$ as shown in Fig. 3. This suggests that anisotropic effects in the entrance  channel of the Li+CaH reaction play a minor role. For these initial states, a minimal basis set with $N_\text{max}=2$ was used. In contrast, the small capture probabilities of spin-aligned reactants tend to be highly sensitive to the value of $N_\text{max}$,  making it necessary to employ much larger rotational basis sets with $N_\text{max}=55$. The large rotational basis sets are required to account for the large anisotropy of the Li-CaH interaction, as shown in our previous work on nonreactive spin relaxation in ultracold Li-CaH collisions \cite{Tscherbul:11}.


\begin{thebibliography}{0}
\expandafter\ifx\csname natexlab\endcsname\relax\def\natexlab#1{#1}\fi
\expandafter\ifx\csname bibnamefont\endcsname\relax
  \def\bibnamefont#1{#1}\fi
\expandafter\ifx\csname bibfnamefont\endcsname\relax
  \def\bibfnamefont#1{#1}\fi
\expandafter\ifx\csname citenamefont\endcsname\relax
  \def\citenamefont#1{#1}\fi
\expandafter\ifx\csname url\endcsname\relax
  \def\url#1{\texttt{#1}}\fi
\expandafter\ifx\csname urlprefix\endcsname\relax\def\urlprefix{URL }\fi
\providecommand{\bibinfo}[2]{#2}
\providecommand{\eprint}[2][]{\url{#2}}

\end{thebibliography}


\begin{thebibliography}{65}
\expandafter\ifx\csname natexlab\endcsname\relax\def\natexlab#1{#1}\fi
\expandafter\ifx\csname bibnamefont\endcsname\relax
  \def\bibnamefont#1{#1}\fi
\expandafter\ifx\csname bibfnamefont\endcsname\relax
  \def\bibfnamefont#1{#1}\fi
\expandafter\ifx\csname citenamefont\endcsname\relax
  \def\citenamefont#1{#1}\fi
\expandafter\ifx\csname url\endcsname\relax
  \def\url#1{\texttt{#1}}\fi
\expandafter\ifx\csname urlprefix\endcsname\relax\def\urlprefix{URL }\fi
\providecommand{\bibinfo}[2]{#2}
\providecommand{\eprint}[2][]{\url{#2}}

\bibitem[{\citenamefont{Zare}(1998)}]{Zare:98}
\bibinfo{author}{\bibfnamefont{R.~N.} \bibnamefont{Zare}},
  \bibinfo{journal}{Science} \textbf{\bibinfo{volume}{279}},
  \bibinfo{pages}{1875} (\bibinfo{year}{1998}).

\bibitem[{\citenamefont{Shapiro and Brumer}(2012)}]{Shapiro:12}
\bibinfo{author}{\bibfnamefont{M.}~\bibnamefont{Shapiro}} \bibnamefont{and}
  \bibinfo{author}{\bibfnamefont{P.}~\bibnamefont{Brumer}},
  \emph{\bibinfo{title}{{\it Quantum Control of Molecular Processes}}}
  (\bibinfo{publisher}{Wiley-VCH, Weinheim}, \bibinfo{year}{2012}).

\bibitem[{\citenamefont{Herschbach}(2006)}]{Herschbach:06}
\bibinfo{author}{\bibfnamefont{D.}~\bibnamefont{Herschbach}},
  \bibinfo{journal}{Eur. Phys. J. D} \textbf{\bibinfo{volume}{38}},
  \bibinfo{pages}{3} (\bibinfo{year}{2006}).

\bibitem[{\citenamefont{Perreault et~al.}(2017)\citenamefont{Perreault,
  Mukherjee, and Zare}}]{Perreault:17}
\bibinfo{author}{\bibfnamefont{W.~E.} \bibnamefont{Perreault}},
  \bibinfo{author}{\bibfnamefont{N.}~\bibnamefont{Mukherjee}},
  \bibnamefont{and} \bibinfo{author}{\bibfnamefont{R.~N.} \bibnamefont{Zare}},
  \bibinfo{journal}{Science} \textbf{\bibinfo{volume}{358}},
  \bibinfo{pages}{356} (\bibinfo{year}{2017}).

\bibitem[{\citenamefont{Perreault et~al.}(2018)\citenamefont{Perreault,
  Mukherjee, and Zare}}]{Perreault:18}
\bibinfo{author}{\bibfnamefont{W.~E.} \bibnamefont{Perreault}},
  \bibinfo{author}{\bibfnamefont{N.}~\bibnamefont{Mukherjee}},
  \bibnamefont{and} \bibinfo{author}{\bibfnamefont{R.~N.} \bibnamefont{Zare}},
  \bibinfo{journal}{Nature Chemistry} \textbf{\bibinfo{volume}{10}},
  \bibinfo{pages}{561} (\bibinfo{year}{2018}).

\bibitem[{\citenamefont{Krems}(2008)}]{Krems:08}
\bibinfo{author}{\bibfnamefont{R.~V.} \bibnamefont{Krems}},
  \bibinfo{journal}{Phys. Chem. Chem. Phys.} \textbf{\bibinfo{volume}{10}},
  \bibinfo{pages}{4079} (\bibinfo{year}{2008}).

\bibitem[{\citenamefont{Balakrishnan}(2016)}]{Balakrishnan:16}
\bibinfo{author}{\bibfnamefont{N.}~\bibnamefont{Balakrishnan}},
  \bibinfo{journal}{J. Chem. Phys.} \textbf{\bibinfo{volume}{145}},
  \bibinfo{pages}{150901} (\bibinfo{year}{2016}).

\bibitem[{\citenamefont{Bohn et~al.}(2017)\citenamefont{Bohn, Rey, and
  Ye}}]{Bohn:17}
\bibinfo{author}{\bibfnamefont{J.~L.} \bibnamefont{Bohn}},
  \bibinfo{author}{\bibfnamefont{A.~M.} \bibnamefont{Rey}}, \bibnamefont{and}
  \bibinfo{author}{\bibfnamefont{J.}~\bibnamefont{Ye}},
  \bibinfo{journal}{Science} \textbf{\bibinfo{volume}{357}},
  \bibinfo{pages}{1002} (\bibinfo{year}{2017}).

\bibitem[{\citenamefont{Lemeshko et~al.}(2013)\citenamefont{Lemeshko, Krems,
  Doyle, and Kais}}]{Lemeshko:13}
\bibinfo{author}{\bibfnamefont{M.}~\bibnamefont{Lemeshko}},
  \bibinfo{author}{\bibfnamefont{R.~V.} \bibnamefont{Krems}},
  \bibinfo{author}{\bibfnamefont{J.~M.} \bibnamefont{Doyle}}, \bibnamefont{and}
  \bibinfo{author}{\bibfnamefont{S.}~\bibnamefont{Kais}},
  \bibinfo{journal}{Mol. Phys} \textbf{\bibinfo{volume}{111}},
  \bibinfo{pages}{1648} (\bibinfo{year}{2013}).

\bibitem[{\citenamefont{Klein et~al.}(2016)\citenamefont{Klein, Shagam,
  Skomorowski, {\.Z}uchowski, Pawlak, Janssen, Moiseyev, van~de Meerakker,
  van~der Avoird, Koch et~al.}}]{Klein:16}
\bibinfo{author}{\bibfnamefont{A.}~\bibnamefont{Klein}},
  \bibinfo{author}{\bibfnamefont{Y.}~\bibnamefont{Shagam}},
  \bibinfo{author}{\bibfnamefont{W.}~\bibnamefont{Skomorowski}},
  \bibinfo{author}{\bibfnamefont{P.~S.} \bibnamefont{{\.Z}uchowski}},
  \bibinfo{author}{\bibfnamefont{M.}~\bibnamefont{Pawlak}},
  \bibinfo{author}{\bibfnamefont{L.~M.~C.} \bibnamefont{Janssen}},
  \bibinfo{author}{\bibfnamefont{N.}~\bibnamefont{Moiseyev}},
  \bibinfo{author}{\bibfnamefont{S.~Y.~T.} \bibnamefont{van~de Meerakker}},
  \bibinfo{author}{\bibfnamefont{A.}~\bibnamefont{van~der Avoird}},
  \bibinfo{author}{\bibfnamefont{C.~P.} \bibnamefont{Koch}},
  \bibnamefont{et~al.}, \bibinfo{journal}{Nat. Phys.}
  \textbf{\bibinfo{volume}{13}}, \bibinfo{pages}{35} (\bibinfo{year}{2016}).

\bibitem[{\citenamefont{Balakrishnan and Dalgarno}(2001)}]{Balakrishnan:01}
\bibinfo{author}{\bibfnamefont{N.}~\bibnamefont{Balakrishnan}}
  \bibnamefont{and} \bibinfo{author}{\bibfnamefont{A.}~\bibnamefont{Dalgarno}},
  \bibinfo{journal}{Chem. Phys. Lett.} \textbf{\bibinfo{volume}{341}},
  \bibinfo{pages}{652} (\bibinfo{year}{2001}).

\bibitem[{\citenamefont{Park et~al.}(2017)\citenamefont{Park, Yan, Loh, Will,
  and Zwierlein}}]{Park:17}
\bibinfo{author}{\bibfnamefont{J.~W.} \bibnamefont{Park}},
  \bibinfo{author}{\bibfnamefont{Z.~Z.} \bibnamefont{Yan}},
  \bibinfo{author}{\bibfnamefont{H.}~\bibnamefont{Loh}},
  \bibinfo{author}{\bibfnamefont{S.~A.} \bibnamefont{Will}}, \bibnamefont{and}
  \bibinfo{author}{\bibfnamefont{M.~W.} \bibnamefont{Zwierlein}},
  \bibinfo{journal}{Science} \textbf{\bibinfo{volume}{357}},
  \bibinfo{pages}{372} (\bibinfo{year}{2017}).

\bibitem[{\citenamefont{Blackmore et~al.}(2018)\citenamefont{Blackmore,
  Caldwell, Gregory, Bridge, Sawant, Aldegunde, Mur-Petit, Jaksch, Hutson,
  Sauer et~al.}}]{Blackmore:18}
\bibinfo{author}{\bibfnamefont{J.~A.} \bibnamefont{Blackmore}},
  \bibinfo{author}{\bibfnamefont{L.}~\bibnamefont{Caldwell}},
  \bibinfo{author}{\bibfnamefont{P.~D.} \bibnamefont{Gregory}},
  \bibinfo{author}{\bibfnamefont{E.~M.} \bibnamefont{Bridge}},
  \bibinfo{author}{\bibfnamefont{R.}~\bibnamefont{Sawant}},
  \bibinfo{author}{\bibfnamefont{J.}~\bibnamefont{Aldegunde}},
  \bibinfo{author}{\bibfnamefont{J.}~\bibnamefont{Mur-Petit}},
  \bibinfo{author}{\bibfnamefont{D.}~\bibnamefont{Jaksch}},
  \bibinfo{author}{\bibfnamefont{J.~M.} \bibnamefont{Hutson}},
  \bibinfo{author}{\bibfnamefont{B.~E.} \bibnamefont{Sauer}},
  \bibnamefont{et~al.}, \bibinfo{journal}{Quantum Sci. Technol.}
  \textbf{\bibinfo{volume}{4}}, \bibinfo{pages}{014010} (\bibinfo{year}{2018}).

\bibitem[{\citenamefont{Vogels et~al.}(2015)\citenamefont{Vogels, Onvlee,
  Chefdeville, van~der Avoird, Groenenboom, and van~de Meerakker}}]{Vogels:15}
\bibinfo{author}{\bibfnamefont{S.~N.} \bibnamefont{Vogels}},
  \bibinfo{author}{\bibfnamefont{J.}~\bibnamefont{Onvlee}},
  \bibinfo{author}{\bibfnamefont{S.}~\bibnamefont{Chefdeville}},
  \bibinfo{author}{\bibfnamefont{A.}~\bibnamefont{van~der Avoird}},
  \bibinfo{author}{\bibfnamefont{G.~C.} \bibnamefont{Groenenboom}},
  \bibnamefont{and} \bibinfo{author}{\bibfnamefont{S.~Y.~T.}
  \bibnamefont{van~de Meerakker}}, \bibinfo{journal}{Science}
  \textbf{\bibinfo{volume}{350}}, \bibinfo{pages}{787} (\bibinfo{year}{2015}).

\bibitem[{\citenamefont{Vogels et~al.}(2018)\citenamefont{Vogels, Karman,
  K{\l}os, Besemer, Onvlee, van~der Avoird, Groenenboom, and van~de
  Meerakker}}]{Vogels:18}
\bibinfo{author}{\bibfnamefont{S.~N.} \bibnamefont{Vogels}},
  \bibinfo{author}{\bibfnamefont{T.}~\bibnamefont{Karman}},
  \bibinfo{author}{\bibfnamefont{J.}~\bibnamefont{K{\l}os}},
  \bibinfo{author}{\bibfnamefont{M.}~\bibnamefont{Besemer}},
  \bibinfo{author}{\bibfnamefont{J.}~\bibnamefont{Onvlee}},
  \bibinfo{author}{\bibfnamefont{A.}~\bibnamefont{van~der Avoird}},
  \bibinfo{author}{\bibfnamefont{G.~C.} \bibnamefont{Groenenboom}},
  \bibnamefont{and} \bibinfo{author}{\bibfnamefont{S.~Y.~T.}
  \bibnamefont{van~de Meerakker}}, \bibinfo{journal}{Nat. Chem.}
  \textbf{\bibinfo{volume}{10}}, \bibinfo{pages}{435} (\bibinfo{year}{2018}).

\bibitem[{\citenamefont{Ni et~al.}(2010)\citenamefont{Ni, Ospelkaus, Wang,
  Qu{\'e}m{\'e}ner, Neyenhuis, de~Miranda, Bohn, Ye, and Jin}}]{Ni:10}
\bibinfo{author}{\bibfnamefont{K.-K.} \bibnamefont{Ni}},
  \bibinfo{author}{\bibfnamefont{S.}~\bibnamefont{Ospelkaus}},
  \bibinfo{author}{\bibfnamefont{D.}~\bibnamefont{Wang}},
  \bibinfo{author}{\bibfnamefont{G.}~\bibnamefont{Qu{\'e}m{\'e}ner}},
  \bibinfo{author}{\bibfnamefont{B.}~\bibnamefont{Neyenhuis}},
  \bibinfo{author}{\bibfnamefont{M.~H.~G.} \bibnamefont{de~Miranda}},
  \bibinfo{author}{\bibfnamefont{J.~L.} \bibnamefont{Bohn}},
  \bibinfo{author}{\bibfnamefont{J.}~\bibnamefont{Ye}}, \bibnamefont{and}
  \bibinfo{author}{\bibfnamefont{D.~S.} \bibnamefont{Jin}},
  \bibinfo{journal}{Nature (London)} \textbf{\bibinfo{volume}{464}},
  \bibinfo{pages}{1324} (\bibinfo{year}{2010}).

\bibitem[{\citenamefont{Ye et~al.}(2018)\citenamefont{Ye, Guo,
  Gonz{\'a}lez-Mart{\'\i}nez, Qu{\'e}m{\'e}ner, and Wang}}]{Ye:18}
\bibinfo{author}{\bibfnamefont{X.}~\bibnamefont{Ye}},
  \bibinfo{author}{\bibfnamefont{M.}~\bibnamefont{Guo}},
  \bibinfo{author}{\bibfnamefont{M.~L.}
  \bibnamefont{Gonz{\'a}lez-Mart{\'\i}nez}},
  \bibinfo{author}{\bibfnamefont{G.}~\bibnamefont{Qu{\'e}m{\'e}ner}},
  \bibnamefont{and} \bibinfo{author}{\bibfnamefont{D.}~\bibnamefont{Wang}},
  \bibinfo{journal}{Science Advances} \textbf{\bibinfo{volume}{4}},
  \bibinfo{pages}{eaaq0083} (\bibinfo{year}{2018}).

\bibitem[{\citenamefont{Yang et~al.}(2019)\citenamefont{Yang, Zhang, Liu, Liu,
  Nan, Zhao, and Pan}}]{Yang:19}
\bibinfo{author}{\bibfnamefont{H.}~\bibnamefont{Yang}},
  \bibinfo{author}{\bibfnamefont{D.-C.} \bibnamefont{Zhang}},
  \bibinfo{author}{\bibfnamefont{L.}~\bibnamefont{Liu}},
  \bibinfo{author}{\bibfnamefont{Y.-X.} \bibnamefont{Liu}},
  \bibinfo{author}{\bibfnamefont{J.}~\bibnamefont{Nan}},
  \bibinfo{author}{\bibfnamefont{B.}~\bibnamefont{Zhao}}, \bibnamefont{and}
  \bibinfo{author}{\bibfnamefont{J.-W.} \bibnamefont{Pan}},
  \bibinfo{journal}{Science} \textbf{\bibinfo{volume}{363}},
  \bibinfo{pages}{261} (\bibinfo{year}{2019}).

\bibitem[{\citenamefont{Ospelkaus et~al.}(2010)\citenamefont{Ospelkaus, Ni,
  Wang, de~Miranda, Neyenhuis, Qu{\'e}m{\'e}ner, Julienne, Bohn, Jin, and
  Ye}}]{Ospelkaus:10}
\bibinfo{author}{\bibfnamefont{S.}~\bibnamefont{Ospelkaus}},
  \bibinfo{author}{\bibfnamefont{K.-K.} \bibnamefont{Ni}},
  \bibinfo{author}{\bibfnamefont{D.}~\bibnamefont{Wang}},
  \bibinfo{author}{\bibfnamefont{M.~H.~G.} \bibnamefont{de~Miranda}},
  \bibinfo{author}{\bibfnamefont{B.}~\bibnamefont{Neyenhuis}},
  \bibinfo{author}{\bibfnamefont{G.}~\bibnamefont{Qu{\'e}m{\'e}ner}},
  \bibinfo{author}{\bibfnamefont{P.~S.} \bibnamefont{Julienne}},
  \bibinfo{author}{\bibfnamefont{J.~L.} \bibnamefont{Bohn}},
  \bibinfo{author}{\bibfnamefont{D.~S.} \bibnamefont{Jin}}, \bibnamefont{and}
  \bibinfo{author}{\bibfnamefont{J.}~\bibnamefont{Ye}},
  \bibinfo{journal}{Science} \textbf{\bibinfo{volume}{327}},
  \bibinfo{pages}{853} (\bibinfo{year}{2010}).

\bibitem[{\citenamefont{de~Miranda et~al.}(2011)\citenamefont{de~Miranda,
  Chotia, Neyenhuis, Wang, Qu{\'e}m{\'e}ner, Ospelkaus, Bohn, Ye, and
  Jin}}]{Miranda:11}
\bibinfo{author}{\bibfnamefont{M.~H.~G.} \bibnamefont{de~Miranda}},
  \bibinfo{author}{\bibfnamefont{A.}~\bibnamefont{Chotia}},
  \bibinfo{author}{\bibfnamefont{B.}~\bibnamefont{Neyenhuis}},
  \bibinfo{author}{\bibfnamefont{D.}~\bibnamefont{Wang}},
  \bibinfo{author}{\bibfnamefont{G.}~\bibnamefont{Qu{\'e}m{\'e}ner}},
  \bibinfo{author}{\bibfnamefont{S.}~\bibnamefont{Ospelkaus}},
  \bibinfo{author}{\bibfnamefont{J.~L.} \bibnamefont{Bohn}},
  \bibinfo{author}{\bibfnamefont{J.}~\bibnamefont{Ye}}, \bibnamefont{and}
  \bibinfo{author}{\bibfnamefont{D.~S.} \bibnamefont{Jin}},
  \bibinfo{journal}{Nat. Phys.} \textbf{\bibinfo{volume}{7}},
  \bibinfo{pages}{502} (\bibinfo{year}{2011}).

\bibitem[{\citenamefont{Shuman et~al.}(2010)\citenamefont{Shuman, Barry, and
  DeMille}}]{Shuman:10}
\bibinfo{author}{\bibfnamefont{E.~S.} \bibnamefont{Shuman}},
  \bibinfo{author}{\bibfnamefont{J.~F.} \bibnamefont{Barry}}, \bibnamefont{and}
  \bibinfo{author}{\bibfnamefont{D.}~\bibnamefont{DeMille}},
  \bibinfo{journal}{Nature (London)} \textbf{\bibinfo{volume}{467}},
  \bibinfo{pages}{820} (\bibinfo{year}{2010}).

\bibitem[{\citenamefont{Barry et~al.}(2014)\citenamefont{Barry, McCarron,
  Norrgard, Steinecker, and DeMille}}]{Barry:14}
\bibinfo{author}{\bibfnamefont{J.~F.} \bibnamefont{Barry}},
  \bibinfo{author}{\bibfnamefont{D.~J.} \bibnamefont{McCarron}},
  \bibinfo{author}{\bibfnamefont{E.~B.} \bibnamefont{Norrgard}},
  \bibinfo{author}{\bibfnamefont{M.~H.} \bibnamefont{Steinecker}},
  \bibnamefont{and} \bibinfo{author}{\bibfnamefont{D.}~\bibnamefont{DeMille}},
  \bibinfo{journal}{Nature (London)} \textbf{\bibinfo{volume}{512}},
  \bibinfo{pages}{286} (\bibinfo{year}{2014}).

\bibitem[{\citenamefont{Truppe et~al.}(2017)\citenamefont{Truppe, Williams,
  Hambach, Caldwell, Fitch, Hinds, Sauer, and Tarbutt}}]{Truppe:17}
\bibinfo{author}{\bibfnamefont{S.}~\bibnamefont{Truppe}},
  \bibinfo{author}{\bibfnamefont{H.~J.} \bibnamefont{Williams}},
  \bibinfo{author}{\bibfnamefont{M.}~\bibnamefont{Hambach}},
  \bibinfo{author}{\bibfnamefont{L.}~\bibnamefont{Caldwell}},
  \bibinfo{author}{\bibfnamefont{N.~J.} \bibnamefont{Fitch}},
  \bibinfo{author}{\bibfnamefont{E.~A.} \bibnamefont{Hinds}},
  \bibinfo{author}{\bibfnamefont{B.~E.} \bibnamefont{Sauer}}, \bibnamefont{and}
  \bibinfo{author}{\bibfnamefont{M.~R.} \bibnamefont{Tarbutt}},
  \bibinfo{journal}{Nat. Phys.} \textbf{\bibinfo{volume}{13}},
  \bibinfo{pages}{1173} (\bibinfo{year}{2017}).

\bibitem[{\citenamefont{Anderegg et~al.}(2017)\citenamefont{Anderegg,
  Augenbraun, Chae, Hemmerling, Hutzler, Ravi, Collopy, Ye, Ketterle, and
  Doyle}}]{Anderegg:17}
\bibinfo{author}{\bibfnamefont{L.}~\bibnamefont{Anderegg}},
  \bibinfo{author}{\bibfnamefont{B.~L.} \bibnamefont{Augenbraun}},
  \bibinfo{author}{\bibfnamefont{E.}~\bibnamefont{Chae}},
  \bibinfo{author}{\bibfnamefont{B.}~\bibnamefont{Hemmerling}},
  \bibinfo{author}{\bibfnamefont{N.~R.} \bibnamefont{Hutzler}},
  \bibinfo{author}{\bibfnamefont{A.}~\bibnamefont{Ravi}},
  \bibinfo{author}{\bibfnamefont{A.}~\bibnamefont{Collopy}},
  \bibinfo{author}{\bibfnamefont{J.}~\bibnamefont{Ye}},
  \bibinfo{author}{\bibfnamefont{W.}~\bibnamefont{Ketterle}}, \bibnamefont{and}
  \bibinfo{author}{\bibfnamefont{J.~M.} \bibnamefont{Doyle}},
  \bibinfo{journal}{Phys. Rev. Lett.} \textbf{\bibinfo{volume}{119}},
  \bibinfo{pages}{103201} (\bibinfo{year}{2017}).

\bibitem[{\citenamefont{Anderegg et~al.}(2018)\citenamefont{Anderegg,
  Augenbraun, Bao, Burchesky, Cheuk, Ketterle, and Doyle}}]{Anderegg:18}
\bibinfo{author}{\bibfnamefont{L.}~\bibnamefont{Anderegg}},
  \bibinfo{author}{\bibfnamefont{B.~L.} \bibnamefont{Augenbraun}},
  \bibinfo{author}{\bibfnamefont{Y.}~\bibnamefont{Bao}},
  \bibinfo{author}{\bibfnamefont{S.}~\bibnamefont{Burchesky}},
  \bibinfo{author}{\bibfnamefont{L.~W.} \bibnamefont{Cheuk}},
  \bibinfo{author}{\bibfnamefont{W.}~\bibnamefont{Ketterle}}, \bibnamefont{and}
  \bibinfo{author}{\bibfnamefont{J.~M.} \bibnamefont{Doyle}},
  \bibinfo{journal}{Nat. Phys.} \textbf{\bibinfo{volume}{14}},
  \bibinfo{pages}{890} (\bibinfo{year}{2018}).

\bibitem[{\citenamefont{Cheuk et~al.}(2018)\citenamefont{Cheuk, Anderegg,
  Augenbraun, Bao, Burchesky, Ketterle, and Doyle}}]{Cheuk:18}
\bibinfo{author}{\bibfnamefont{L.~W.} \bibnamefont{Cheuk}},
  \bibinfo{author}{\bibfnamefont{L.}~\bibnamefont{Anderegg}},
  \bibinfo{author}{\bibfnamefont{B.~L.} \bibnamefont{Augenbraun}},
  \bibinfo{author}{\bibfnamefont{Y.}~\bibnamefont{Bao}},
  \bibinfo{author}{\bibfnamefont{S.}~\bibnamefont{Burchesky}},
  \bibinfo{author}{\bibfnamefont{W.}~\bibnamefont{Ketterle}}, \bibnamefont{and}
  \bibinfo{author}{\bibfnamefont{J.~M.} \bibnamefont{Doyle}},
  \bibinfo{journal}{Phys. Rev. Lett.} \textbf{\bibinfo{volume}{121}},
  \bibinfo{pages}{083201} (\bibinfo{year}{2018}).

\bibitem[{\citenamefont{McCarron et~al.}(2018)\citenamefont{McCarron,
  Steinecker, Zhu, and DeMille}}]{McCarron:18}
\bibinfo{author}{\bibfnamefont{D.~J.} \bibnamefont{McCarron}},
  \bibinfo{author}{\bibfnamefont{M.~H.} \bibnamefont{Steinecker}},
  \bibinfo{author}{\bibfnamefont{Y.}~\bibnamefont{Zhu}}, \bibnamefont{and}
  \bibinfo{author}{\bibfnamefont{D.}~\bibnamefont{DeMille}},
  \bibinfo{journal}{Phys. Rev. Lett.} \textbf{\bibinfo{volume}{121}},
  \bibinfo{pages}{013202} (\bibinfo{year}{2018}).

\bibitem[{\citenamefont{Lim et~al.}(2018)\citenamefont{Lim, Almond, Trigatzis,
  Devlin, Fitch, Sauer, Tarbutt, and Hinds}}]{Lim:18}
\bibinfo{author}{\bibfnamefont{J.}~\bibnamefont{Lim}},
  \bibinfo{author}{\bibfnamefont{J.~R.} \bibnamefont{Almond}},
  \bibinfo{author}{\bibfnamefont{M.~A.} \bibnamefont{Trigatzis}},
  \bibinfo{author}{\bibfnamefont{J.~A.} \bibnamefont{Devlin}},
  \bibinfo{author}{\bibfnamefont{N.~J.} \bibnamefont{Fitch}},
  \bibinfo{author}{\bibfnamefont{B.~E.} \bibnamefont{Sauer}},
  \bibinfo{author}{\bibfnamefont{M.~R.} \bibnamefont{Tarbutt}},
  \bibnamefont{and} \bibinfo{author}{\bibfnamefont{E.~A.} \bibnamefont{Hinds}},
  \bibinfo{journal}{Phys. Rev. Lett.} \textbf{\bibinfo{volume}{120}},
  \bibinfo{pages}{123201} (\bibinfo{year}{2018}).

\bibitem[{\citenamefont{Kozyryev et~al.}(2017)\citenamefont{Kozyryev, Baum,
  Matsuda, Augenbraun, Anderegg, Sedlack, and Doyle}}]{Kozyryev:17}
\bibinfo{author}{\bibfnamefont{I.}~\bibnamefont{Kozyryev}},
  \bibinfo{author}{\bibfnamefont{L.}~\bibnamefont{Baum}},
  \bibinfo{author}{\bibfnamefont{K.}~\bibnamefont{Matsuda}},
  \bibinfo{author}{\bibfnamefont{B.~L.} \bibnamefont{Augenbraun}},
  \bibinfo{author}{\bibfnamefont{L.}~\bibnamefont{Anderegg}},
  \bibinfo{author}{\bibfnamefont{A.~P.} \bibnamefont{Sedlack}},
  \bibnamefont{and} \bibinfo{author}{\bibfnamefont{J.~M.} \bibnamefont{Doyle}},
  \bibinfo{journal}{Phys. Rev. Lett.} \textbf{\bibinfo{volume}{118}},
  \bibinfo{pages}{173201} (\bibinfo{year}{2017}).

\bibitem[{\citenamefont{Tscherbul and Krems}(2006)}]{Tscherbul:06}
\bibinfo{author}{\bibfnamefont{T.~V.} \bibnamefont{Tscherbul}}
  \bibnamefont{and} \bibinfo{author}{\bibfnamefont{R.~V.} \bibnamefont{Krems}},
  \bibinfo{journal}{Phys. Rev. Lett.} \textbf{\bibinfo{volume}{97}},
  \bibinfo{pages}{083201} (\bibinfo{year}{2006}).

\bibitem[{\citenamefont{Abrahamsson et~al.}(2007)\citenamefont{Abrahamsson,
  Tscherbul, and Krems}}]{Abrahamsson:07}
\bibinfo{author}{\bibfnamefont{E.}~\bibnamefont{Abrahamsson}},
  \bibinfo{author}{\bibfnamefont{T.~V.} \bibnamefont{Tscherbul}},
  \bibnamefont{and} \bibinfo{author}{\bibfnamefont{R.~V.} \bibnamefont{Krems}},
  \bibinfo{journal}{J. Chem. Phys.} \textbf{\bibinfo{volume}{127}},
  \bibinfo{pages}{044302} (\bibinfo{year}{2007}).

\bibitem[{\citenamefont{Knoop et~al.}(2010)\citenamefont{Knoop, Ferlaino,
  Berninger, Mark, N\"agerl, Grimm, D'Incao, and Esry}}]{Knoop:10}
\bibinfo{author}{\bibfnamefont{S.}~\bibnamefont{Knoop}},
  \bibinfo{author}{\bibfnamefont{F.}~\bibnamefont{Ferlaino}},
  \bibinfo{author}{\bibfnamefont{M.}~\bibnamefont{Berninger}},
  \bibinfo{author}{\bibfnamefont{M.}~\bibnamefont{Mark}},
  \bibinfo{author}{\bibfnamefont{H.-C.} \bibnamefont{N\"agerl}},
  \bibinfo{author}{\bibfnamefont{R.}~\bibnamefont{Grimm}},
  \bibinfo{author}{\bibfnamefont{J.~P.} \bibnamefont{D'Incao}},
  \bibnamefont{and} \bibinfo{author}{\bibfnamefont{B.~D.} \bibnamefont{Esry}},
  \bibinfo{journal}{Phys. Rev. Lett.} \textbf{\bibinfo{volume}{104}},
  \bibinfo{pages}{053201} (\bibinfo{year}{2010}).

\bibitem[{\citenamefont{Li and Cong}(2019)}]{Li:19}
\bibinfo{author}{\bibfnamefont{J.-L.} \bibnamefont{Li}} \bibnamefont{and}
  \bibinfo{author}{\bibfnamefont{S.-L.} \bibnamefont{Cong}},
  \bibinfo{journal}{Phys. Rev. A} \textbf{\bibinfo{volume}{99}},
  \bibinfo{pages}{022708} (\bibinfo{year}{2019}).

\bibitem[{\citenamefont{Janssen et~al.}(2013)\citenamefont{Janssen, van~der
  Avoird, and Groenenboom}}]{Janssen:13}
\bibinfo{author}{\bibfnamefont{L.~M.~C.} \bibnamefont{Janssen}},
  \bibinfo{author}{\bibfnamefont{A.}~\bibnamefont{van~der Avoird}},
  \bibnamefont{and} \bibinfo{author}{\bibfnamefont{G.~C.}
  \bibnamefont{Groenenboom}}, \bibinfo{journal}{Phys. Lev. Lett.}
  \textbf{\bibinfo{volume}{110}}, \bibinfo{pages}{063201}
  (\bibinfo{year}{2013}).

\bibitem[{\citenamefont{Rackham
  et~al.}(2003{\natexlab{a}})\citenamefont{Rackham, Huarte-Larranaga, and
  Manolopoulos}}]{Rackham:03}
\bibinfo{author}{\bibfnamefont{E.~J.} \bibnamefont{Rackham}},
  \bibinfo{author}{\bibfnamefont{F.}~\bibnamefont{Huarte-Larranaga}},
  \bibnamefont{and} \bibinfo{author}{\bibfnamefont{D.~E.}
  \bibnamefont{Manolopoulos}}, \bibinfo{journal}{Chem. Phys. Lett.}
  \textbf{\bibinfo{volume}{343}}, \bibinfo{pages}{356}
  (\bibinfo{year}{2003}{\natexlab{a}}).

\bibitem[{\citenamefont{Rackham
  et~al.}(2003{\natexlab{b}})\citenamefont{Rackham, Gonzalez-Lezana, and
  Manolopoulos}}]{Rackham:03a}
\bibinfo{author}{\bibfnamefont{E.~J.} \bibnamefont{Rackham}},
  \bibinfo{author}{\bibfnamefont{T.}~\bibnamefont{Gonzalez-Lezana}},
  \bibnamefont{and} \bibinfo{author}{\bibfnamefont{D.~E.}
  \bibnamefont{Manolopoulos}}, \bibinfo{journal}{J. Chem. Phys.}
  \textbf{\bibinfo{volume}{119}}, \bibinfo{pages}{12895}
  (\bibinfo{year}{2003}{\natexlab{b}}).

\bibitem[{\citenamefont{Alexander et~al.}(2004)\citenamefont{Alexander,
  Rackham, and Manolopoulos}}]{Alexander:04}
\bibinfo{author}{\bibfnamefont{M.~H.} \bibnamefont{Alexander}},
  \bibinfo{author}{\bibfnamefont{E.~J.} \bibnamefont{Rackham}},
  \bibnamefont{and} \bibinfo{author}{\bibfnamefont{D.~E.}
  \bibnamefont{Manolopoulos}}, \bibinfo{journal}{J. Chem. Phys.}
  \textbf{\bibinfo{volume}{121}}, \bibinfo{pages}{5221} (\bibinfo{year}{2004}).

\bibitem[{\citenamefont{Gonz\'alez-Lezana}(2007)}]{GonzalezLezana:07}
\bibinfo{author}{\bibfnamefont{T.}~\bibnamefont{Gonz\'alez-Lezana}},
  \bibinfo{journal}{Int. Rev. Phys. Chem.} \textbf{\bibinfo{volume}{26}},
  \bibinfo{pages}{29} (\bibinfo{year}{2007}).

\bibitem[{\citenamefont{Miller}(1970)}]{Miller:70}
\bibinfo{author}{\bibfnamefont{W.}~\bibnamefont{Miller}}, \bibinfo{journal}{J.
  Chem. Phys.} \textbf{\bibinfo{volume}{52}}, \bibinfo{pages}{543}
  (\bibinfo{year}{1970}).

\bibitem[{\citenamefont{Clary}(1990)}]{Clary:90}
\bibinfo{author}{\bibfnamefont{D.~C.} \bibnamefont{Clary}},
  \bibinfo{journal}{Annu. Rev. Phys. Chem.} \textbf{\bibinfo{volume}{41}},
  \bibinfo{pages}{61} (\bibinfo{year}{1990}).

\bibitem[{\citenamefont{Qu\'em\'ener and Bohn}(2010)}]{Quemener:10}
\bibinfo{author}{\bibfnamefont{G.}~\bibnamefont{Qu\'em\'ener}}
  \bibnamefont{and} \bibinfo{author}{\bibfnamefont{J.~L.} \bibnamefont{Bohn}},
  \bibinfo{journal}{Phys. Rev. A} \textbf{\bibinfo{volume}{81}},
  \bibinfo{pages}{022702} (\bibinfo{year}{2010}).

\bibitem[{\citenamefont{Gonz\'alez-Mart\'inez
  et~al.}(2014)\citenamefont{Gonz\'alez-Mart\'inez, Dulieu, Larr\'egaray, and
  Bonnet}}]{GonzalezMartinez:14}
\bibinfo{author}{\bibfnamefont{M.~L.~T.} \bibnamefont{Gonz\'alez-Mart\'inez}},
  \bibinfo{author}{\bibfnamefont{O.}~\bibnamefont{Dulieu}},
  \bibinfo{author}{\bibfnamefont{P.}~\bibnamefont{Larr\'egaray}},
  \bibnamefont{and} \bibinfo{author}{\bibfnamefont{L.}~\bibnamefont{Bonnet}},
  \bibinfo{journal}{Phys. Rev. A} \textbf{\bibinfo{volume}{90}},
  \bibinfo{pages}{052716} (\bibinfo{year}{2014}).

\bibitem[{\citenamefont{Idziaszek and Julienne}(2010)}]{Idziaszek:10}
\bibinfo{author}{\bibfnamefont{Z.}~\bibnamefont{Idziaszek}} \bibnamefont{and}
  \bibinfo{author}{\bibfnamefont{P.~S.} \bibnamefont{Julienne}},
  \bibinfo{journal}{Phys. Rev. Lett.} \textbf{\bibinfo{volume}{104}},
  \bibinfo{pages}{113202} (\bibinfo{year}{2010}).

\bibitem[{\citenamefont{Gao}(2010)}]{Gao:10}
\bibinfo{author}{\bibfnamefont{B.}~\bibnamefont{Gao}}, \bibinfo{journal}{Phys.
  Rev. Lett.} \textbf{\bibinfo{volume}{105}}, \bibinfo{pages}{263203}
  (\bibinfo{year}{2010}).

\bibitem[{\citenamefont{Qu{\'e}m{\'e}ner and Julienne}(2012)}]{Quemener:12}
\bibinfo{author}{\bibfnamefont{G.}~\bibnamefont{Qu{\'e}m{\'e}ner}}
  \bibnamefont{and} \bibinfo{author}{\bibfnamefont{P.~S.}
  \bibnamefont{Julienne}}, \bibinfo{journal}{Chem. Rev.}
  \textbf{\bibinfo{volume}{112}}, \bibinfo{pages}{4949} (\bibinfo{year}{2012}).

\bibitem[{\citenamefont{Dagdigian}(2017)}]{Dagdigian:17}
\bibinfo{author}{\bibfnamefont{P.~J.} \bibnamefont{Dagdigian}},
  \bibinfo{journal}{J. Chem. Phys.} \textbf{\bibinfo{volume}{146}},
  \bibinfo{pages}{224308} (\bibinfo{year}{2017}).

\bibitem[{\citenamefont{Croft et~al.}(2017)\citenamefont{Croft, Makrides, Li,
  Petrov, Kendrick, Balakrishnan, and Kotochigova}}]{Croft:17}
\bibinfo{author}{\bibfnamefont{J.~F.~E.} \bibnamefont{Croft}},
  \bibinfo{author}{\bibfnamefont{C.}~\bibnamefont{Makrides}},
  \bibinfo{author}{\bibfnamefont{M.}~\bibnamefont{Li}},
  \bibinfo{author}{\bibfnamefont{A.}~\bibnamefont{Petrov}},
  \bibinfo{author}{\bibfnamefont{B.~K.} \bibnamefont{Kendrick}},
  \bibinfo{author}{\bibfnamefont{N.}~\bibnamefont{Balakrishnan}},
  \bibnamefont{and}
  \bibinfo{author}{\bibfnamefont{S.}~\bibnamefont{Kotochigova}},
  \bibinfo{journal}{Nat. Commun.} \textbf{\bibinfo{volume}{8}},
  \bibinfo{pages}{15897} (\bibinfo{year}{2017}).

\bibitem[{\citenamefont{Makrides et~al.}(2015)\citenamefont{Makrides, Hazra,
  Pradhan, Petrov, Kendrick, Gonz\'alez-Lezana, Balakrishnan, and
  Kotochigova}}]{Makrides:15}
\bibinfo{author}{\bibfnamefont{C.}~\bibnamefont{Makrides}},
  \bibinfo{author}{\bibfnamefont{J.}~\bibnamefont{Hazra}},
  \bibinfo{author}{\bibfnamefont{G.~B.} \bibnamefont{Pradhan}},
  \bibinfo{author}{\bibfnamefont{A.}~\bibnamefont{Petrov}},
  \bibinfo{author}{\bibfnamefont{B.~K.} \bibnamefont{Kendrick}},
  \bibinfo{author}{\bibfnamefont{T.}~\bibnamefont{Gonz\'alez-Lezana}},
  \bibinfo{author}{\bibfnamefont{N.}~\bibnamefont{Balakrishnan}},
  \bibnamefont{and}
  \bibinfo{author}{\bibfnamefont{S.}~\bibnamefont{Kotochigova}},
  \bibinfo{journal}{Phys. Rev. A} \textbf{\bibinfo{volume}{91}},
  \bibinfo{pages}{012708} (\bibinfo{year}{2015}).

\bibitem[{\citenamefont{Tscherbul and Buchachenko}(2015)}]{Tscherbul:15a}
\bibinfo{author}{\bibfnamefont{T.~V.} \bibnamefont{Tscherbul}}
  \bibnamefont{and} \bibinfo{author}{\bibfnamefont{A.~A.}
  \bibnamefont{Buchachenko}}, \bibinfo{journal}{New J. Phys.}
  \textbf{\bibinfo{volume}{17}}, \bibinfo{pages}{035010}
  (\bibinfo{year}{2015}).

\bibitem[{\citenamefont{Singh et~al.}(2012)\citenamefont{Singh, Hardman, Tariq,
  Lu, Ellis, Morrison, and Weinstein}}]{Singh:12}
\bibinfo{author}{\bibfnamefont{V.}~\bibnamefont{Singh}},
  \bibinfo{author}{\bibfnamefont{K.~S.} \bibnamefont{Hardman}},
  \bibinfo{author}{\bibfnamefont{N.}~\bibnamefont{Tariq}},
  \bibinfo{author}{\bibfnamefont{M.-J.} \bibnamefont{Lu}},
  \bibinfo{author}{\bibfnamefont{A.}~\bibnamefont{Ellis}},
  \bibinfo{author}{\bibfnamefont{M.~J.} \bibnamefont{Morrison}},
  \bibnamefont{and} \bibinfo{author}{\bibfnamefont{J.~D.}
  \bibnamefont{Weinstein}}, \bibinfo{journal}{Phys. Lev. Lett.}
  \textbf{\bibinfo{volume}{108}}, \bibinfo{pages}{203201}
  (\bibinfo{year}{2012}).

\bibitem[{\citenamefont{Carr et~al.}(2009)\citenamefont{Carr, DeMille, Krems,
  and Ye}}]{Carr:09}
\bibinfo{author}{\bibfnamefont{L.~D.} \bibnamefont{Carr}},
  \bibinfo{author}{\bibfnamefont{D.}~\bibnamefont{DeMille}},
  \bibinfo{author}{\bibfnamefont{R.~V.} \bibnamefont{Krems}}, \bibnamefont{and}
  \bibinfo{author}{\bibfnamefont{J.}~\bibnamefont{Ye}}, \bibinfo{journal}{New
  J. Phys} \textbf{\bibinfo{volume}{11}}, \bibinfo{pages}{055049}
  (\bibinfo{year}{2009}).

\bibitem[{\citenamefont{Tscherbul et~al.}(2011)\citenamefont{Tscherbul,
  K{\l}os, and Buchachenko}}]{Tscherbul:11}
\bibinfo{author}{\bibfnamefont{T.~V.} \bibnamefont{Tscherbul}},
  \bibinfo{author}{\bibfnamefont{J.}~\bibnamefont{K{\l}os}}, \bibnamefont{and}
  \bibinfo{author}{\bibfnamefont{A.~A.} \bibnamefont{Buchachenko}},
  \bibinfo{journal}{Phys. Rev. A} \textbf{\bibinfo{volume}{84}},
  \bibinfo{pages}{040701(R)} (\bibinfo{year}{2011}).

\bibitem[{\citenamefont{Lim et~al.}(2015)\citenamefont{Lim, Frye, Hutson, and
  Tarbutt}}]{Lim:15}
\bibinfo{author}{\bibfnamefont{J.}~\bibnamefont{Lim}},
  \bibinfo{author}{\bibfnamefont{M.~D.} \bibnamefont{Frye}},
  \bibinfo{author}{\bibfnamefont{J.~M.} \bibnamefont{Hutson}},
  \bibnamefont{and} \bibinfo{author}{\bibfnamefont{M.~R.}
  \bibnamefont{Tarbutt}}, \bibinfo{journal}{Phys. Rev. A}
  \textbf{\bibinfo{volume}{92}}, \bibinfo{pages}{053419}
  (\bibinfo{year}{2015}).

\bibitem[{\citenamefont{Morita et~al.}(2018)\citenamefont{Morita, Kosicki,
  \ifmmode~\dot{Z}\else \.{Z}\fi{}uchowski, and Tscherbul}}]{Morita:18}
\bibinfo{author}{\bibfnamefont{M.}~\bibnamefont{Morita}},
  \bibinfo{author}{\bibfnamefont{M.~B.} \bibnamefont{Kosicki}},
  \bibinfo{author}{\bibfnamefont{P.~S.} \bibnamefont{\ifmmode~\dot{Z}\else
  \.{Z}\fi{}uchowski}}, \bibnamefont{and} \bibinfo{author}{\bibfnamefont{T.~V.}
  \bibnamefont{Tscherbul}}, \bibinfo{journal}{Phys. Rev. A}
  \textbf{\bibinfo{volume}{98}}, \bibinfo{pages}{042702}
  (\bibinfo{year}{2018}).

\bibitem[{\citenamefont{Tscherbul and Dalgarno}(2010)}]{Tscherbul:10}
\bibinfo{author}{\bibfnamefont{T.~V.} \bibnamefont{Tscherbul}}
  \bibnamefont{and} \bibinfo{author}{\bibfnamefont{A.}~\bibnamefont{Dalgarno}},
  \bibinfo{journal}{J. Chem. Phys.} \textbf{\bibinfo{volume}{133}},
  \bibinfo{pages}{184104} (\bibinfo{year}{2010}).

\bibitem[{SM()}]{SM}
\bibinfo{note}{See Supplemental Material at [http://link.aps.org/
  supplemental/] for details of numerical calculations and convergence tests.}

\bibitem[{\citenamefont{Groenenboom and Janssen}(2010)}]{Groenenboom:10}
\bibinfo{author}{\bibfnamefont{G.~C.} \bibnamefont{Groenenboom}}
  \bibnamefont{and} \bibinfo{author}{\bibfnamefont{L.~M.~C.}
  \bibnamefont{Janssen}}, \bibinfo{journal}{in {\it Tutorials in molecular
  reaction dynamics}, ed. by M. Brouard and C. Vallance, RSC, Cambridge}
  (\bibinfo{year}{2010}).

\bibitem[{\citenamefont{Moerdijk et~al.}(1996)\citenamefont{Moerdijk, Verhaar,
  and Nagtegaal}}]{Moerdijk:96}
\bibinfo{author}{\bibfnamefont{A.~J.} \bibnamefont{Moerdijk}},
  \bibinfo{author}{\bibfnamefont{B.~J.} \bibnamefont{Verhaar}},
  \bibnamefont{and} \bibinfo{author}{\bibfnamefont{T.~M.}
  \bibnamefont{Nagtegaal}}, \bibinfo{journal}{Phys. Rev. A}
  \textbf{\bibinfo{volume}{53}}, \bibinfo{pages}{4343} (\bibinfo{year}{1996}).

\bibitem[{\citenamefont{Sikorsky et~al.}(2018)\citenamefont{Sikorsky, Morita,
  Meir, Buchachenko, Ben-shlomi, Akerman, Narevicius, Tscherbul, and
  Ozeri}}]{Sikorsky:18}
\bibinfo{author}{\bibfnamefont{T.}~\bibnamefont{Sikorsky}},
  \bibinfo{author}{\bibfnamefont{M.}~\bibnamefont{Morita}},
  \bibinfo{author}{\bibfnamefont{Z.}~\bibnamefont{Meir}},
  \bibinfo{author}{\bibfnamefont{A.~A.} \bibnamefont{Buchachenko}},
  \bibinfo{author}{\bibfnamefont{R.}~\bibnamefont{Ben-shlomi}},
  \bibinfo{author}{\bibfnamefont{N.}~\bibnamefont{Akerman}},
  \bibinfo{author}{\bibfnamefont{E.}~\bibnamefont{Narevicius}},
  \bibinfo{author}{\bibfnamefont{T.~V.} \bibnamefont{Tscherbul}},
  \bibnamefont{and} \bibinfo{author}{\bibfnamefont{R.}~\bibnamefont{Ozeri}},
  \bibinfo{journal}{Phys. Rev. Lett.} \textbf{\bibinfo{volume}{121}},
  \bibinfo{pages}{173402} (\bibinfo{year}{2018}).

\bibitem[{\citenamefont{Warehime and K\l{}os}(2015)}]{Warehime:15}
\bibinfo{author}{\bibfnamefont{M.}~\bibnamefont{Warehime}} \bibnamefont{and}
  \bibinfo{author}{\bibfnamefont{J.}~\bibnamefont{K\l{}os}},
  \bibinfo{journal}{Phys. Rev. A} \textbf{\bibinfo{volume}{92}},
  \bibinfo{pages}{032703} (\bibinfo{year}{2015}).

\bibitem[{\citenamefont{Puri et~al.}(2017)\citenamefont{Puri, Mills, Schneider,
  Simbotin, Montgomery, C{\^o}t{\'e}, Suits, and Hudson}}]{Puri:17}
\bibinfo{author}{\bibfnamefont{P.}~\bibnamefont{Puri}},
  \bibinfo{author}{\bibfnamefont{M.}~\bibnamefont{Mills}},
  \bibinfo{author}{\bibfnamefont{C.}~\bibnamefont{Schneider}},
  \bibinfo{author}{\bibfnamefont{I.}~\bibnamefont{Simbotin}},
  \bibinfo{author}{\bibfnamefont{J.~A.} \bibnamefont{Montgomery}},
  \bibinfo{author}{\bibfnamefont{R.}~\bibnamefont{C{\^o}t{\'e}}},
  \bibinfo{author}{\bibfnamefont{A.~G.} \bibnamefont{Suits}}, \bibnamefont{and}
  \bibinfo{author}{\bibfnamefont{E.~R.} \bibnamefont{Hudson}},
  \bibinfo{journal}{Science} \textbf{\bibinfo{volume}{357}},
  \bibinfo{pages}{1370} (\bibinfo{year}{2017}).

\bibitem[{\citenamefont{Yang et~al.}(2018)\citenamefont{Yang, Li, Chen, Xie,
  Suits, Campbell, Guo, and Hudson}}]{Yang:18}
\bibinfo{author}{\bibfnamefont{T.}~\bibnamefont{Yang}},
  \bibinfo{author}{\bibfnamefont{A.}~\bibnamefont{Li}},
  \bibinfo{author}{\bibfnamefont{G.~K.} \bibnamefont{Chen}},
  \bibinfo{author}{\bibfnamefont{C.}~\bibnamefont{Xie}},
  \bibinfo{author}{\bibfnamefont{A.~G.} \bibnamefont{Suits}},
  \bibinfo{author}{\bibfnamefont{W.~C.} \bibnamefont{Campbell}},
  \bibinfo{author}{\bibfnamefont{H.}~\bibnamefont{Guo}}, \bibnamefont{and}
  \bibinfo{author}{\bibfnamefont{E.~R.} \bibnamefont{Hudson}},
  \bibinfo{journal}{J. Phys. Chem. Lett.} \textbf{\bibinfo{volume}{9}},
  \bibinfo{pages}{3555} (\bibinfo{year}{2018}).

\bibitem[{\citenamefont{Zhang and Willitsch}(2017)}]{Zhang:17}
\bibinfo{author}{\bibfnamefont{D.}~\bibnamefont{Zhang}} \bibnamefont{and}
  \bibinfo{author}{\bibfnamefont{S.}~\bibnamefont{Willitsch}},
  \bibinfo{journal}{in {\it Cold Chemistry: Molecular Scattering and Reactivity
  Near Absolute Zero}, ed. by O. Dulieu and A. Osterwalder, RSC, Cambridge}
  (\bibinfo{year}{2017}).

\bibitem[{\citenamefont{Hall and Willitsch}(2012)}]{Hall:12}
\bibinfo{author}{\bibfnamefont{F.~H.~J.} \bibnamefont{Hall}} \bibnamefont{and}
  \bibinfo{author}{\bibfnamefont{S.}~\bibnamefont{Willitsch}},
  \bibinfo{journal}{Phys. Rev. Lett.} \textbf{\bibinfo{volume}{109}},
  \bibinfo{pages}{233202} (\bibinfo{year}{2012}).

\bibitem[{\citenamefont{Meyer and Bohn}(2011)}]{Meyer:11}
\bibinfo{author}{\bibfnamefont{E.~R.} \bibnamefont{Meyer}} \bibnamefont{and}
  \bibinfo{author}{\bibfnamefont{J.~L.} \bibnamefont{Bohn}},
  \bibinfo{journal}{Phys. Rev. A} \textbf{\bibinfo{volume}{83}},
  \bibinfo{pages}{032714} (\bibinfo{year}{2011}).

\end{thebibliography}

\begin{thebibliography}{15}%
\makeatletter
\providecommand \@ifxundefined [1]{%
 \@ifx{#1\undefined}
}%
\providecommand \@ifnum [1]{%
 \ifnum #1\expandafter \@firstoftwo
 \else \expandafter \@secondoftwo
 \fi
}%
\providecommand \@ifx [1]{%
 \ifx #1\expandafter \@firstoftwo
 \else \expandafter \@secondoftwo
 \fi
}%
\providecommand \natexlab [1]{#1}%
\providecommand \enquote  [1]{``#1''}%
\providecommand \bibnamefont  [1]{#1}%
\providecommand \bibfnamefont [1]{#1}%
\providecommand \citenamefont [1]{#1}%
\providecommand \href@noop [0]{\@secondoftwo}%
\providecommand \href [0]{\begingroup \@sanitize@url \@href}%
\providecommand \@href[1]{\@@startlink{#1}\@@href}%
\providecommand \@@href[1]{\endgroup#1\@@endlink}%
\providecommand \@sanitize@url [0]{\catcode `\\12\catcode `\$12\catcode
  `\&12\catcode `\#12\catcode `\^12\catcode `\_12\catcode `\%12\relax}%
\providecommand \@@startlink[1]{}%
\providecommand \@@endlink[0]{}%
\providecommand \url  [0]{\begingroup\@sanitize@url \@url }%
\providecommand \@url [1]{\endgroup\@href {#1}{\urlprefix }}%
\providecommand \urlprefix  [0]{URL }%
\providecommand \Eprint [0]{\href }%
\providecommand \doibase [0]{http://dx.doi.org/}%
\providecommand \selectlanguage [0]{\@gobble}%
\providecommand \bibinfo  [0]{\@secondoftwo}%
\providecommand \bibfield  [0]{\@secondoftwo}%
\providecommand \translation [1]{[#1]}%
\providecommand \BibitemOpen [0]{}%
\providecommand \bibitemStop [0]{}%
\providecommand \bibitemNoStop [0]{.\EOS\space}%
\providecommand \EOS [0]{\spacefactor3000\relax}%
\providecommand \BibitemShut  [1]{\csname bibitem#1\endcsname}%
\let\auto@bib@innerbib\@empty
\bibitem [{\citenamefont {Rackham}\ \emph
  {et~al.}(2003{\natexlab{a}})\citenamefont {Rackham}, \citenamefont
  {Huarte-Larranaga},\ and\ \citenamefont {Manolopoulos}}]{Rackham:03}%
  \BibitemOpen
  \bibfield  {author} {\bibinfo {author} {\bibfnamefont {E.~J.}\ \bibnamefont
  {Rackham}}, \bibinfo {author} {\bibfnamefont {F.}~\bibnamefont
  {Huarte-Larranaga}}, \ and\ \bibinfo {author} {\bibfnamefont {D.~E.}\
  \bibnamefont {Manolopoulos}},\ }\href@noop {} {\bibfield  {journal} {\bibinfo
   {journal} {Chem. Phys. Lett.}\ }\textbf {\bibinfo {volume} {343}},\ \bibinfo
  {pages} {356} (\bibinfo {year} {2003}{\natexlab{a}})}\BibitemShut {NoStop}%
\bibitem [{\citenamefont {Rackham}\ \emph
  {et~al.}(2003{\natexlab{b}})\citenamefont {Rackham}, \citenamefont
  {Gonzalez-Lezana},\ and\ \citenamefont {Manolopoulos}}]{Rackham:03a}%
  \BibitemOpen
  \bibfield  {author} {\bibinfo {author} {\bibfnamefont {E.~J.}\ \bibnamefont
  {Rackham}}, \bibinfo {author} {\bibfnamefont {T.}~\bibnamefont
  {Gonzalez-Lezana}}, \ and\ \bibinfo {author} {\bibfnamefont {D.~E.}\
  \bibnamefont {Manolopoulos}},\ }\href@noop {} {\bibfield  {journal} {\bibinfo
   {journal} {J. Chem. Phys.}\ }\textbf {\bibinfo {volume} {119}},\ \bibinfo
  {pages} {12895} (\bibinfo {year} {2003}{\natexlab{b}})}\BibitemShut {NoStop}%
\bibitem [{\citenamefont {Alexander}\ \emph {et~al.}(2004)\citenamefont
  {Alexander}, \citenamefont {Rackham},\ and\ \citenamefont
  {Manolopoulos}}]{Alexander:04}%
  \BibitemOpen
  \bibfield  {author} {\bibinfo {author} {\bibfnamefont {M.~H.}\ \bibnamefont
  {Alexander}}, \bibinfo {author} {\bibfnamefont {E.~J.}\ \bibnamefont
  {Rackham}}, \ and\ \bibinfo {author} {\bibfnamefont {D.~E.}\ \bibnamefont
  {Manolopoulos}},\ }\href@noop {} {\bibfield  {journal} {\bibinfo  {journal}
  {J. Chem. Phys.}\ }\textbf {\bibinfo {volume} {121}},\ \bibinfo {pages}
  {5221} (\bibinfo {year} {2004})}\BibitemShut {NoStop}%
\bibitem {Tscherbul:10} 
T. V. Tscherbul and  A. Dalgarno, J. Chem. Phys. {\bf 133}, 184104 (2010).
\bibitem [{\citenamefont {Manolopoulos}()}]{Manolopoulos:06}%
  \BibitemOpen
  \bibfield  {author} {\bibinfo {author} {\bibfnamefont {D.~E.}\ \bibnamefont
  {Manolopoulos}},\ }\href@noop {} {\bibinfo  {journal} {{\it Airy function
  capture boundary conditions} (unpublished notes, 2006)}\ }\BibitemShut
  {NoStop}%
\bibitem{Tscherbul:07}
T.~V. Tscherbul, J. K\l{}os, L. Rajchel,  and R. V. Krems, Phys. Rev. A  {\bf 75}, 033416 (2007).
\bibitem [{\citenamefont {Tscherbul}(2018)}]{Tscherbul:18b}%
  \BibitemOpen
  \bibfield  {author} {\bibinfo {author} {\bibfnamefont {T.~V.}\ \bibnamefont
  {Tscherbul}},\ }\enquote {\bibinfo {title} {{\it Cold Chemistry: Molecular
  Scattering and Reactivity Near Absolute Zero}},}\ \ (\bibinfo  {publisher}
  {Royal Society of Chemistry},\ \bibinfo {year} {2018})\ Chap.~\bibinfo
  {chapter} {6}\BibitemShut {NoStop}%
\bibitem [{\citenamefont {Werner}\ \emph {et~al.}(2012)\citenamefont {Werner},
  \citenamefont {Knowles}, \citenamefont {Knizia}, \citenamefont {Manby},\ and\
  \citenamefont {Sch{\"u}tz}}]{molpro}%
  \BibitemOpen
  \bibfield  {author} {\bibinfo {author} {\bibfnamefont {H.-J.}\ \bibnamefont
  {Werner}}, \bibinfo {author} {\bibfnamefont {P.~J.}\ \bibnamefont {Knowles}},
  \bibinfo {author} {\bibfnamefont {G.}~\bibnamefont {Knizia}}, \bibinfo
  {author} {\bibfnamefont {F.~R.}\ \bibnamefont {Manby}}, \ and\ \bibinfo
  {author} {\bibfnamefont {M.}~\bibnamefont {Sch{\"u}tz}},\ }\href@noop {}
  {\bibfield  {journal} {\bibinfo  {journal} {WIREs Comput Mol Sci}\ }\textbf
  {\bibinfo {volume} {2}},\ \bibinfo {pages} {242} (\bibinfo {year}
  {2012})}\BibitemShut {NoStop}%
\bibitem [{\citenamefont {Tscherbul}\ \emph {et~al.}(2011)\citenamefont
  {Tscherbul}, \citenamefont {K{\l}os},\ and\ \citenamefont
  {Buchachenko}}]{Tscherbul:11}%
  \BibitemOpen
  \bibfield  {author} {\bibinfo {author} {\bibfnamefont {T.~V.}\ \bibnamefont
  {Tscherbul}}, \bibinfo {author} {\bibfnamefont {J.}~\bibnamefont {K{\l}os}},
  \ and\ \bibinfo {author} {\bibfnamefont {A.~A.}\ \bibnamefont
  {Buchachenko}},\ }\href@noop {} {\bibfield  {journal} {\bibinfo  {journal}
  {Phys. Rev. A}\ }\textbf {\bibinfo {volume} {84}},\ \bibinfo {pages}
  {040701(R)} (\bibinfo {year} {2011})}\BibitemShut {NoStop}%
\bibitem [{\citenamefont {Knowles}\ \emph {et~al.}(1993)\citenamefont
  {Knowles}, \citenamefont {Hampel},\ and\ \citenamefont
  {Werner}}]{knowles:93}%
  \BibitemOpen
  \bibfield  {author} {\bibinfo {author} {\bibfnamefont {P.~J.}\ \bibnamefont
  {Knowles}}, \bibinfo {author} {\bibfnamefont {C.}~\bibnamefont {Hampel}}, \
  and\ \bibinfo {author} {\bibfnamefont {H.-J.}\ \bibnamefont {Werner}},\
  }\href@noop {} {\bibfield  {journal} {\bibinfo  {journal} {J. Chem. Phys.}\
  }\textbf {\bibinfo {volume} {99}},\ \bibinfo {pages} {5219} (\bibinfo {year}
  {1993})}\BibitemShut {NoStop}%
\bibitem [{\citenamefont {Werner}\ and\ \citenamefont
  {Knowles}(1988)}]{werner:88}%
  \BibitemOpen
  \bibfield  {author} {\bibinfo {author} {\bibfnamefont {H.}~\bibnamefont
  {Werner}}\ and\ \bibinfo {author} {\bibfnamefont {P.~J.}\ \bibnamefont
  {Knowles}},\ }\href@noop {} {\bibfield  {journal} {\bibinfo  {journal} {J.
  Chem. Phys.}\ }\textbf {\bibinfo {volume} {89}},\ \bibinfo {pages} {5803}
  (\bibinfo {year} {1988})}\BibitemShut {NoStop}%
\bibitem [{\citenamefont {Dunning}(1989)}]{dunning:89}%
  \BibitemOpen
  \bibfield  {author} {\bibinfo {author} {\bibfnamefont {T.~H.}\ \bibnamefont
  {Dunning}},\ }\href@noop {} {\bibfield  {journal} {\bibinfo  {journal} {J.
  Chem. Phys.}\ }\textbf {\bibinfo {volume} {90}},\ \bibinfo {pages} {1007}
  (\bibinfo {year} {1989})}\BibitemShut {NoStop}%
\bibitem [{\citenamefont {Koput}\ and\ \citenamefont
  {Peterson}(2002)}]{Koput:2002}%
  \BibitemOpen
  \bibfield  {author} {\bibinfo {author} {\bibfnamefont {J.}~\bibnamefont
  {Koput}}\ and\ \bibinfo {author} {\bibfnamefont {K.~A.}\ \bibnamefont
  {Peterson}},\ }\href@noop {} {\bibfield  {journal} {\bibinfo  {journal} {J.
  Phys. Chem. A}\ }\textbf {\bibinfo {volume} {106}},\ \bibinfo {pages} {9595}
  (\bibinfo {year} {2002})}\BibitemShut {NoStop}%
\bibitem [{\citenamefont {Ho}\ and\ \citenamefont {Rabitz}(1996)}]{ho:96}%
  \BibitemOpen
  \bibfield  {author} {\bibinfo {author} {\bibfnamefont {T.-S.}\ \bibnamefont
  {Ho}}\ and\ \bibinfo {author} {\bibfnamefont {H.}~\bibnamefont {Rabitz}},\
  }\href@noop {} {\bibfield  {journal} {\bibinfo  {journal} {J. Chem. Phys.}\
  }\textbf {\bibinfo {volume} {104}},\ \bibinfo {pages} {2584} (\bibinfo {year}
  {1996})}\BibitemShut {NoStop}%
\bibitem [{\citenamefont {Tscherbul}\ and\ \citenamefont
  {Buchachenko}(2015)}]{Tscherbul:15a}%
  \BibitemOpen
  \bibfield  {author} {\bibinfo {author} {\bibfnamefont {T.~V.}\ \bibnamefont
  {Tscherbul}}\ and\ \bibinfo {author} {\bibfnamefont {A.~A.}\ \bibnamefont
  {Buchachenko}},\ }\href@noop {} {\bibfield  {journal} {\bibinfo  {journal}
  {New J. Phys.}\ }\textbf {\bibinfo {volume} {17}},\ \bibinfo {pages} {035010}
  (\bibinfo {year} {2015})}\BibitemShut {NoStop}%
\end{thebibliography}
\end{document}